\documentclass{aa}  

\usepackage[justification=centering, labelfont=bf]{caption}
\usepackage{graphicx}
\usepackage[utf8]{inputenc}
\usepackage[T1]{fontenc}
\usepackage[english]{babel}
\usepackage{palatino}
\usepackage{txfonts}
\usepackage{natbib}
\usepackage{stfloats}
\bibpunct{(}{)}{;}{a}{}{,}
\usepackage[colorlinks,linkcolor=blue,urlcolor=blue,citecolor=blue]{hyperref}
\usepackage{booktabs}
\usepackage{array}
\usepackage{tabularx}
\usepackage{bm}
\usepackage{float}
\usepackage{multirow}
\usepackage{epstopdf,epsfig}
\usepackage{amsmath}
\usepackage{float}
\usepackage{balance}
\usepackage{placeins}
\usepackage[switch]{lineno}
\floatstyle{plaintop}
\restylefloat{table}
\usepackage[flushleft]{threeparttable}

\makeatletter
\@fpsep\textheight
\makeatother
\newcommand{\quotes}[1]{``#1''}

\begin{document} 

\title{Direct imaging of molten protoplanets in nearby young stellar associations}

\author{
Irene Bonati\inst{1,2}, 
Tim Lichtenberg\inst{2,3}, 
Dan J. Bower\inst{4}, 
Miles L. Timpe\inst{5} 
\and Sascha P. Quanz\inst{6}
}

\institute{
Earth-Life Science Institute, Tokyo Institute of Technology, 2-12-1 Ookayama I7E-312, Meguro-ku, Tokyo 152-8550, Japan\\
\email{irene.bonati@elsi.jp}
\and
Institute of Geophysics, ETH Zurich, Sonneggstrasse 5, 8092 Zurich, Switzerland
\and
Atmospheric, Oceanic and Planetary Physics, University of Oxford, Parks Rd, Oxford OX1 3PU, United Kingdom
\and
Center for Space and Habitability, University of Bern, Gesellschaftsstrasse 6, 3012 Bern, Switzerland     
\and
Institute for Computational Science, University of Zurich, Winterthurerstrasse 190, 8057 Zurich, Switzerland
\and
Institute for Particle Physics and Astrophysics, ETH Zurich, Wolfgang-Pauli-Strasse 27, 8093 Zurich, Switzerland
}

\date{ Received; accepted} 

\authorrunning{Bonati et al.}

\abstract{During their formation and early evolution, rocky planets undergo multiple global melting events due to accretionary collisions with other protoplanets. The detection and characterization of their post-collision afterglows (magma oceans) can yield important clues about the origin and evolution of the solar and extrasolar planet population. Here, we quantitatively assess the observational prospects to detect the radiative signature of forming planets covered by such collision-induced magma oceans in nearby young stellar associations with future direct imaging facilities. We have compared performance estimates for near- and mid-infrared instruments to be installed at ESO's Extremely Large Telescope (ELT), and a potential space-based mission called Large Interferometer for Exoplanets (LIFE). We modelled the frequency and timing of energetic collisions using \textit{N}-body models of planet formation for different stellar types, and determine the cooling of the resulting magma oceans with an insulating atmosphere. We find that the probability of detecting at least one magma ocean planet depends on the observing duration and the distribution of atmospheric properties among rocky protoplanets. However, the prospects for detection significantly increase for young and close stellar targets, which show the highest frequencies of giant impacts. For intensive reconnaissance with a K band (2.2 $\mu m$) ELT filter or a 5.6 $\mu m$ LIFE filter, the $\beta$ Pictoris, Columba, TW Hydrae, and Tucana-Horologium associations represent promising candidates for detecting a molten protoplanet. Our results motivate the exploration of magma ocean planets using the ELT and underline the importance of space-based direct imaging facilities to investigate and characterize planet formation and evolution in the solar vicinity. Direct imaging of magma oceans will advance our understanding of the early interior, surface and atmospheric properties of terrestrial worlds.}


\maketitle

\section{Introduction}
Theoretical studies of planet formation suggest that during their early phase of evolution, rocky planets undergo multiple magma ocean stages as a result of impacts \citep[e.g.,][]{Benz, Melosh90,Tonks}, internal heating from the decay of short-lived radioisotopes \citep{Elkins}, and the release of potential energy during core formation \citep{Flasar,Sasaki}. As an example, the giant impact that formed the Moon \citep{1975Icar...24..504H} melted a large portion of the Earth's mantle to produce a global magma ocean \citep[e.g.,][]{Canup,Naka} that subsequently cooled and crystallized. Processes taking place during magma ocean solidification are of fundamental importance for the subsequent evolution of a planet, as they determine its early thermal and chemical structure, atmospheric composition, tectonic behavior, and ultimately its habitability \citep{Massol,2018SSRv..214...76I}.

Planets covered by a magma ocean have not yet been observed directly. Nevertheless, hot molten bodies are likely common in the Universe; one such example is the tidally-locked super-Earth 55 Cnc $e$.  Heat maps of 55 Cnc $e$ indicate a temperature on the star-facing side that is sufficient to melt silicates \citep{Demory,Hammond17}. Employing direct imaging techniques to newly forming planets could lead to the detection of another type of molten surface: surface magma oceans that are the product of a recent giant impact. Such a detection would provide direct observational constraints for theoretical models of planet formation, interior and atmospheric dynamics, as well as insights into the origin and diversification of planets in the solar system and extrasolar systems.

The composition of the atmosphere plays an important role in the direct detection of hot molten protoplanets.  In general, a primary atmosphere grows by degassing of over-saturated volatiles (e.g., H\textsubscript{2}O, CO\textsubscript{2}, CH\textsubscript{4}, N\textsubscript{2}) during the crystallization of a magma ocean. The accumulation of greenhouse gases within the pre-existing (primordial) atmosphere exerts a thermal blanketing effect which inhibits heat radiation to space and hence slows down the cooling of the interior relative to no atmosphere being present \citep{Matsui2,ET08,Lebrun,Hamanonat}. The composition and evolution of the outgassed atmosphere thus strongly controls the longevity and observability of molten planets.

On the one hand, prolonged cooling timescales \citep[up to 100 Myr, see][]{Hamanonat} increase the probability of detection because a planet remains hotter (and hence brighter) for longer. On the other hand, dense and optically thick atmospheres can hinder the direct observation of planetary surfaces.  This is because the observed (brightness) temperature of a planet, as dictated by the convective-radiative equilibrium of its atmosphere, may be lower than its actual surface temperature. Hence magma ocean bodies may appear less bright and thus be more challenging to detect.

Following a giant impact, the molten surface of a hot protoplanet is expected to be between 1000 K and 2000 K and thus emits primarily at infrared (IR) wavelengths \citep{Stern94,Zhang03,Mamajek07}. As a magma ocean cools it reaches the so-called \quotes{rheological transition} around a critical melt fraction of 40\%; at this transition the viscosity of the solid--melt mixture abruptly increases by orders of magnitude \citep{Solomatov,Costa} and the magma ocean behaves rheologically as a solid. Past studies have addressed the detectability of magma ocean planets by predicting the wavelengths that are able to penetrate thick outgassed atmospheres. \citet{MillerRicci} explored thermal emission spectra of magma ocean bodies for a range of atmospheric compositions, showing that thermal radiation can pass through a series of atmospheric near-IR windows and potentially be observed by future telescopes. \citet{Lupu} further investigated high-resolution spectra for systems composed of atmospheres in equilibrium with magmas of different compositions, and concluded that wavelengths of 1-4 $\mu$m are the most favorable for the direct detection of such bodies.

Spectral features are expected to evolve during magma ocean cooling due to the evolution of the surface and atmosphere of the hot protoplanets.  In particular, the atmosphere is replenished by volatiles outgassing from the interior, and eroded by hydrodynamic escape caused by extreme ultraviolet (XUV) radiation emitted by the young host star. \citet{Hamanonat,Hamano} linked the thermal evolution of magma ocean planets to their spectral variations, and found that bodies located beyond a critical orbital distance $a_{\mathrm{cr}}$ ($\sim 0.8$ AU for a solar-type star), so-called type I planets, solidify fast and display thermal emissions that decay within several million years. Most of the water acquired during planetary formation is preserved and forms primordial water oceans. Conversely, on type II planets, located inside the critical orbital distance $a_{\mathrm{cr}}$, magma oceans can be maintained for much longer---perhaps more than 100 Myr.  The timescale depends on the initial water inventory and the hydrodynamic escape process, which progressively leads to the desiccation of the planet. Bodies with extended solidification times emit significant thermal radiation. Based on planet-to-star contrast estimates, \citet{Hamano} found that the $K_{\mathrm{s}}$ (2.16 $\mu$m) and $L$ bands (3.55 $\mu$m) are the most favorable for detection.

With the advent of a number of imminent direct imaging facilities and potential future missions, a comprehensive and quantitative assessment of the detectability of bodies in their early formation stage is desired. Past investigations have focused on magma ocean evolution and its associated spectra without providing clear prospects for future detectability based on telescope specifications. These studies also typically include an advanced description of energy transfer in Earth-like atmospheres at the expense of a more complete model of interior dynamics. 

Here, we aim to quantify the likelihood of observing magma ocean planets by convolving the expected occurrence rate of giant impacts with the frequency and ages of close-by stellar associations in order to find the best targets and telescope specifications for future observations. We use \textit{N}-body simulations \citep{Genga} to constrain the occurrence rate and timing of energetic collisions inducing global magma oceans during terrestrial planet formation around different stellar types. Subsequently, we chart the evolution of the surface temperature of these molten protoplanets using an interior model devised to describe the cooling and crystallization history of rocky bodies \citep{Bowerm}. The detectability of these bodies is then assessed by employing performance estimates for instruments that are to be installed at ESO's 39-m Extremely Large Telescope (ELT) and a hypothetical space-based mid-infrared interferometer \citep{Kammerer,2018arXiv180706088Q,2018arXiv180709996D}. Finally, combining the results of the aforementioned modeling steps enables us to compute the likelihood of detecting at least one magma ocean planet within nearby young stellar associations.

\section{Methods}
\label{sec:methods}

\subsection{Detectability assessment} 
\label{sec:detect}
The ability of future telescopes to detect protoplanetary collisional afterglows is assessed using technical specifications or performance estimates for ELT instruments and a potential space telescope, called Large Interferometer for Exoplanets (LIFE\footnote{\href{https://www.life-space-mission.com}{https://www.life-space-mission.com}}). LIFE is inspired by the Darwin mission concept, which was originally proposed to ESA in 2007 \citep{Leger} with the goal of directly detecting and characterizing extrasolar planets at mid-infrared wavelengths using a free-flying interferometer in space \citep{Darwin}. However, Darwin, and similarly NASA's TPF-I concept, was never realized due to technical challenges and uncertainties related to the expected scientific yield. The growing interest in the search for extrasolar life and significantly improved yield estimates based on results from NASA's Kepler spacecraft \citep[e.g.,][]{Winn15} motivate a reconsideration of similar space missions for future explorations \citep{Kammerer,2018arXiv180709996D,2018arXiv180706088Q}. 

Young stellar associations represent promising candidates for harboring forming planetary systems subject to energetic encounters between planetary bodies. We consider ten young stellar groups in the solar vicinity (Table~\ref{tab:SA}) that could be searched for magma ocean planets in the future.
\setlength{\tabcolsep}{4pt}   
\begin{table}[ht]
  \centering
  \begin{threeparttable}
		\begin{tabular}
        {c c c c c c c}
			\toprule
            \toprule
			\multicolumn{6}{r}{\textbf{No. stars}}\\ 
			\cmidrule(l){4-6}
			\textbf{Association} & \textbf{\textit{d} (pc) }&\textbf{ Age (Myr)}&\textbf{A}&\textbf{ G} & \textbf{M}& \textbf{Ref.}\\
			\midrule 
			AB Doradus & 20 & 150 & 0 & 23 &8 & 3\\ 
			$\beta$ Pictoris\textsuperscript{a} & 37\textsuperscript{a} & 23 & 6 & 16 & 9 &1\textsuperscript{a}\\ 
			$\beta$ Tucanae & 43 & 45 & 4 & 0 &1 & 4\\ 
			Tucana Horologium & 48 & 45 & 2& 27 &5&2,5\\
			Columba & 50 & 42& 2 & 37 &1 & 2\\ 
			TW Hydrae & 53 & 10 & 2 & 0 &33 & 6\\ 
			Carina & 65 & 45 & 0 & 22 &0 & 2\\ 
			32 Orionis & 92 & 22 & 2 & 7 &2 & 7,8\\ 
			$\eta$ Chamaeleontis & 94 & 11 & 1 & 0 &11& 2\\ 
			$\chi$1 For & 99 & 50 & 2 & 0 &0 & 9\\ 
			\bottomrule 
		\end{tabular}
		\captionsetup{justification=justified,font=small}
		\caption[Parameters of young stellar associations]{Parameters of the young stellar associations considered for the detectability assessment.}
		\label{tab:SA}
    \begin{tablenotes}
      \small
      \item \textbf{Notes}: The distances (\textit{d}) and ages of the stellar associations are from \citet{Mamajek} and references therein. The references in the last column give the number of A-, G-, and M-star members listed: (1) \citet{Beta}, (2) \citet{Torres}, (3) \citet{ABDoradus}, (4) \citet{Simbad}, (5) \citet{Kraus}, (6) \citet{Gagne}, (7) \citet{Bell2}, (8) \citet{Sh}, (9) \citet{Mamajek}. We consider only bona fide stellar members to be conservative. \textsuperscript{a}For the $\beta$ Pictoris moving group we computed the mean distances from a combined list of Table 1 and 2 of \cite{Beta}, and took the spectral types from the SIMBAD database where they were not provided. The mean distance of 15 pc given in \citet{Mamajek} seems to be a typo. 
    \end{tablenotes}
  \end{threeparttable}
\end{table}
We assume that all stars belonging to a given stellar association have the same age as the association and have planets forming around them. The number of G-type members in Table \ref{tab:SA} also comprises F- and early K-type stars (up to K4) for simplicity.  

We calculate the probability $P_{\mathrm{MO}}$ to detect at least one magma ocean planet in a given set of observed stars in a stellar association using
\begin{linenomath*}
\begin{align} \label{eq:master_equation_42}
\resizebox{0.93\hsize}{!}{$P_{\mathrm{MO}}(\lambda_{\mathrm{cen}},\mathrm{d},\tau_\mathrm{{*}},\epsilon)
=1-\prod_{i=1}^{i=n_{*}}\left(1-\frac{\bar{n}_{\mathrm{GI,i}}\cdot \Delta t_{\mathrm{MO_{\mathrm{\bar{R}}},i}}}{\Delta t_{\mathrm{bin,i}}}\right),$}
\end{align}
\end{linenomath*}
\noindent
which implicitly depends on $\lambda_{\mathrm{cen}}$, the central wavelength of the considered filter, the distance $d$ to the stellar association, the age of the stellar sample $\tau_{*}$, and the assumed planetary atmospheric emissivity $\epsilon$. $n_{\mathrm{*}}$ is the number of stars of a given spectral type (A, G or M) located inside a given stellar association, according to Table \ref{tab:SA}. $\bar{n}_{\mathrm{GI,i}}$  is the number of detectable giant impacts within a specified time interval of planet formation ($\Delta t_{\mathrm{bin,i}}=20$ Myr), which accounts for the age of the considered stellar association. $\Delta t_{\mathrm{MO_{\mathrm{\bar{R}}},i}}$ indicates the time interval within which a magma ocean planet of radius $\bar{R}$ is bright enough to be directly observed by future instruments. $P_{\mathrm{MO}}$ depends on $\tau_{*}$ through $\bar{n}_{\mathrm{GI,i}}$ (depending on the age of a stellar association, and hence the stage of planet formation, planetary objects will experience a different number of magma ocean inducing giant impacts), and on $\epsilon$ through $\Delta t_{\mathrm{MO_{\mathrm{\bar{R}}},i}}$ (depending on the thickness of its atmosphere, a magma ocean planet will display a more or less strong radiative signature). $\bar{n}_{\mathrm{GI,i}}$ and $\Delta t_{\mathrm{MO_{\mathrm{\bar{R}}},i}}$ are determined from \textit{N}-body simulations (Sect. \ref{sec:nbody}) and a magma ocean evolution model (Sect. \ref{sec:MO}), respectively. Both parameters also depend on the imaging filter wavelength ($\lambda_{\mathrm{cen}}$) and corresponding telescope performance estimates (Table \ref{tab:Perf}), as well as on the distance of the targeted stellar association ($d$). 

The telescope parameters most relevant for the detectability assessment are the angular resolution and the sensitivity (Table \ref{tab:Perf}). A successful imaging program requires both high angular resolution and high sensitivity. In this study we do not explicitly consider contrast-limited observations, but comment on the relevance of contrast performance in Sect. \ref{sec:likelihood}. The limiting angular resolution of a single-aperture telescope is defined by its inner working angle (IWA), that is, the smallest angular distance still allowing a clear spatial separation of the planet from its host star. For the ELT, we assume an IWA of $2\lambda_\mathrm{cen}/D$, where $\lambda_\mathrm{cen}$ is the central wavelength of the filter and $D$ is the aperture size of the telescope. For LIFE, a free-flying nulling interferometer, we assume that a nulling baseline of up to $B = 168$ m can be accommodated \citep{Leger,Darwin}, resulting theoretically in a best possible IWA of $\lambda_\mathrm{cen}/(4B) = 3$ mas at a wavelength of 10 $\mu$m. To be conservative, the IWAs assumed in our analysis for LIFE are almost a factor of three larger than this estimate. The sensitivity quantifies the faintest signal that a telescope can detect at a certain wavelength within a specified time interval. Increasing the observation time reduces the noise of the thermal background emission and hence enables the detection of progressively fainter bodies. We assume that background-noise limited performance can be achieved at separations as small as the assumed IWAs. 

\renewcommand{\arraystretch}{1.15} 	
\setlength{\tabcolsep}{16pt} 
\begin{table}[bt]
  \begin{threeparttable}
		\centering 
		\begin{tabular}{c c c c} 
			\toprule
            \toprule
			Filter &$\lambda_{\mathrm{cen}}$ & IWA & 	  Sensitivity\textsuperscript{a}\\
              & ($\mu$m)& (mas) & ($\mu$Jy)\\
			\midrule 
			\textbf{\underline{ELT}}\\
            K&2.2&24&0.008\\
			L&3.8&42&0.263\\
			N&11.6&128&17.530\\
			\textbf{\underline{LIFE}}\\
			F560W&5.6&5&0.056\\
			F1000W&10&8&0.188\\
			\bottomrule
		\end{tabular}
	\captionsetup{justification=justified,font=small
	}
		\caption[Filter and performance estimates of the ELT and LIFE]{Filter and performance estimates of ELT and LIFE.} 
		\label{tab:Perf}
    \begin{tablenotes}
      \small
      \item \textbf{Notes}: 
      We assume that the sensitivity limits are achieved down to separations as small as the listed inner working angles (IWA). \textsuperscript{a} Sensitivities are given as 5-$\sigma$ detection limits in 5 hours of on-source integration time. It is assumed that sensitivities scale inversely with the square root of the integration time. The ELT K band sensitivity is based on the ELT simulator (\url{https://www.eso.org/observing/etc/}) and assumes a pixel scale of 5 mas/px, a laser tomography adaptive optics (LTAO) system and an airmass of 1.2. The sensitivities for ELT L and N band are updates to the values for the METIS instrument \citep[METIS consortium, private communication; cf.][]{Quanz}. The sensitivities for LIFE are based on the values presented in \citet[][]{Kammerer}. More detailed investigations, also concerning the final contrast performance of the various instruments, are required in the future to refine the detection limits.
    \end{tablenotes}
  \end{threeparttable}
\end{table}

For the ELT we consider two filters from the Mid-infrared ELT Imager and Spectrograph \citep[METIS,][]{Quanz2,brandl2018}, one in the L band and one in the N band, with central wavelengths $\lambda_{\mathrm{cen}}$ of 3.8 $\mu$m and 11.6 $\mu$m, respectively. In addition, we perform our analysis for a generic instrument operating in the K band filter ($\lambda_{\mathrm{cen}} = 2.2$ $\mu$m). For the LIFE telescope, we consider two filters, F560W and F1000W, having central wavelengths $\lambda_{\mathrm{cen}}$ of 5.6 $\mu$m and 10 $\mu$m, respectively. The assumed sensitivities for LIFE are based on the faint source detection limits of the Mid-Infrared Instrument (MIRI) installed on the James Webb Space Telescope (JWST), but take into account the reduced throughput of a nulling interferometer \citep[cf.][]{James, Kammerer}.

In our analysis, a planet is considered as detectable if its angular separation $\theta = \arctan
{a/d}$ (with $a$ being the planet's semi-major axis, and $d$ the distance to the stellar system) exceeds the instrument’s IWA, and if its flux observed in a specific band is higher than the instrument’s sensitivity. Inclination effects are ignored by implicitly assuming face-on orbits. The comparison of angular separation and observed flux with the assumptions for IWA and sensitivity of the considered instruments (Table \ref{tab:Perf}) enables the determination of parameters related to the detectable bodies, such as their size and brightness temperature.

\subsection{Timing and energy of giant impacts}
\label{sec:nbody}
We model the giant impact stage of planet formation using the \textit{N}-body code GENGA \citep[Gravitational ENcounters with Gpu Acceleration,][]{Genga}. GENGA is a hybrid symplectic \textit{N}-body integrator running fully on GPUs and is specifically designed for simulations of planet formation and orbital evolution. We investigate the occurrence rate of magma ocean-inducing impacts by simulating planet formation around A- (2 \(M_\odot\)), G- (1 \(M_\odot\)), and M-type (0.5 \(M_\odot\)) stars during 200 Myr of evolution, focusing on the giant impact phase. Due to the stochastic nature of \textit{N}-body models \citep{Kokubo,Hoffmann17} we perform five simulations for each stellar type. The initial conditions (Table \ref{tab:ic}) differ between each stellar mass, but are identical for all simulations of a given star type. The initial disk masses and surface density profiles are based on terrestrial planet formation around low-mass stars \citep{Raymond}.

\setlength{\tabcolsep}{3.2pt} 
	\begin{table}[htb] 
		\centering 
		\begin{tabular}{c c c c c c}
			\toprule
            \toprule
			$\textbf{Spec. type}$ & $\boldsymbol{M_{\mathrm{*}}}$ (\(M_\odot\)) & $\boldsymbol{R_{\mathrm{disk}} (AU)}$ & $\boldsymbol{M_{\mathrm{disk}}}$ ($\boldsymbol{M_{\mathrm{\oplus}})}$ & $\boldsymbol{N_{\mathrm{planetesimals}}}$\\ 
			\midrule 
			A & 2.0 & 0.5-5 & 10.0 & 60\\ 
			G & 1.0 & 0.5-4 & 4.95 & 75\\ 
			M & 0.5 & 0.1-1 & 0.70 & 110\\ 
            \bottomrule
		\end{tabular}
		\captionsetup{justification=justified,
font=small}
		\caption[Initial conditions of \textit{N}-body simulations]{Initial conditions of the \textit{N}-body simulations for A-, G-, and M-stars. $M_{\mathrm{*}}$ is the stellar mass. $R_{\mathrm{disk}}$ and $M_{\mathrm{disk}}$ indicate the modeled disk section and the disk mass, respectively. The initial number of bodies ($N_{\mathrm{planetesimals}}$) is constrained to focus on the giant impact phase of planet formation and to ensure reasonable compute time.}
		\label{tab:ic}
\end{table}

The initial planetesimal disk around each star contains a number of planetesimals $N_{\mathrm{planetesimals}}$ with total mass $M_{\rm disk}$ distributed within a circumstellar annulus $R_{\mathrm{disk}}$. The planetesimals are distributed according to a surface density profile $\Sigma \propto r^{-1}$. During the subsequent orbital evolution, any body passing within 0.1 AU or more than 50 AU from the central star is removed from the \textit{N}-body integration. In the former case the body is considered to have collided with the central star, whereas in the latter case it is considered ejected. During the subsequent integration, planetesimals and protoplanets grow larger through pairwise collisions between bodies. In our simulations, colliding bodies undergo perfect accretion (i.e., disruptive and partial accretion events are not considered), with the mass of the resulting body corresponding to the sum of the masses of the colliding planetesimals.

To quantify which of these collisions are sufficiently energetic to form magma oceans, we calculate the specific impact energy $Q$ in the center of mass frame as defined by \cite{LeinhardtStewartI}. The specific impact energy can be interpreted as the kinetic energy of the impact distributed throughout the mass of the resulting body,
\begin{linenomath*}
\begin{align}
	Q = \frac{\mu~v_{\mathrm{imp}}^2}{2 \left( m_{\mathrm{imp}} + m_{\mathrm{tar}} \right)},
\end{align}
\end{linenomath*}
\noindent where $m_{\mathrm{imp}}$ is the mass of the impactor, $m_{\mathrm{tar}}$ is mass of the target, $v_{\mathrm{imp}}$ is impact speed, and $\mu$ is reduced mass,
\begin{linenomath*}
\begin{align}
	\mu = \frac{m_{\mathrm{imp}}~m_{\mathrm{\mathrm{tar}}}}{\left( m_{\mathrm{imp}} + m_{\mathrm{tar}} \right)}.
\end{align}
\end{linenomath*}
We set the critical specific impact energy $Q_{\mathrm{crit}} = 0.3 \times 10^{6}$ J/kg, above which a magma ocean is formed. This value is consistent with energy estimates of the canonical Moon-forming giant impact \citep{Canup}. Regarding the frequency of giant impacts, \citet{Quintana} consider a giant impact energy threshold $Q_{\mathrm{{crit}}}= 2 \times 10^{6}$ J/kg, corresponding to the energy needed to strip half of a planet's atmosphere. Here we consider a smaller value, since atmospheric loss is not needed to generate a magma ocean. We calculate $Q$ for every collision in each simulation to determine the occurrence rate of giant impacts and the radii of the post-impact bodies around each stellar mass. The radii of the post-impact bodies are determined using a mass-radius relationship for rocky planets \citep{Mass}
\begin{linenomath*}
\begin{align} \label{eq:pythagoras3}
\frac{R}{R_{\mathrm{\oplus}}}=1.0154 \left( \frac{M}{M_{\mathrm{\oplus}}} \right)^{1/3.7}, 
\end{align}
\end{linenomath*}
\noindent where $R$ is the planetary radius, $R_\oplus$ is Earth's radius (6371 km), $M$ is planetary mass, and $M_\mathrm{\oplus} = 6\times 10^{24}$ kg is Earth's mass. The magma ocean lifetimes of the post-impact bodies are then calculated using an interior evolution model (Sect. \ref{sec:MO}).

\subsection{Lifetimes of magma oceans}
\label{sec:MO}
We investigate the thermal evolution of magma ocean bodies using an energy balance model with self-consistent thermodynamics of melt and solid silicate phases \citep{Bowerm,Wolf}. This model provides an improved description of the interior dynamics of magma oceans, which are often simplified in previous studies that focus mainly on atmospheric evolution \citep{Hamanonat,Lebrun,Hamano,Tanton_Moon,Lupu}.
It includes a newly-developed equation of state that provides the thermo-physical properties of silicate melt throughout the pressure-temperature range of Earth's mantle \citep{Wolf}. The interior evolution model assumes 1-D spherically symmetric geometry and accounts for heat transfer by conduction, convection, mixing (latent heat transport), and gravitational separation.  We consider bottom-up crystallization where solids first form at the base of the mantle and crystallization proceeds towards the surface.

The rate at which a magma ocean can transfer energy through its outgassed atmosphere constitutes the rate-limiting factor controlling the cooling timescale of the molten body. We consider two different models for the efficiency of energy transport in the atmosphere: (1) a greybody, which parameterizes the insulating effect of the atmosphere using an effective emissivity $\epsilon$ \citep[e.g.,][]{ET08}, and (2) a steam atmosphere parameterization \citep{Zahnle88,Sol93}.
Moreover, our models include a parameterization of the temperature drop that occurs in the ultra-thin ($\lesssim$ 2 cm thickness) thermal boundary layer at the surface of the magma ocean \citep{Abe2}.  The temperature drop scales as the surface temperature cubed \citep[e.g.,][]{Reese}.

We define the lifetime of a magma ocean as the time required for the surface of an initially molten planet to reach the rheological transition \citep[around 40\% melt fraction;][]{Solomatov}.  Beyond this time it is unclear if a cold lid forms at the surface or if the surface remains partially molten due to the delivery of heat by the convecting magma below. Furthermore, a cool surface leads to a cool upper atmosphere that is expected to harbor thick water clouds for H\textsubscript{2}O-CO\textsubscript{2} atmospheres, thus efficiently masking the surface \citep{Marcq17}. Hence our definition of magma ocean lifetime is designed as a minimum duration to give a conservative estimate of the potential for detectability. Other studies typically define the lifetime as the time taken for the planet to become dominantly ($\sim$ 98\%) solid \citep[e.g.,][]{ET08}.

We compute surface temperature evolutions and magma ocean cooling timescales for planetary radii $R \in$ [0.1, 0.3, 0.5, 0.7, 1] $\times$ $R_{\oplus}$ and atmospheric emissivities $\epsilon \in$ [0.01, $\sqrt{10^{-3}}$, 0.1, $\sqrt{10^{-1}}$, 0.5, 1] for the gray atmosphere model. For an Earth-like planet with a 1 bar atmosphere and absorption coefficients of H\textsubscript{2}O of 0.01 m\textsuperscript{2}/kg \citep{Yamamoto} and CO\textsubscript{2} of 0.05 m\textsuperscript{2}/kg \citep{Pujol}, an emissivity $\epsilon=0.01$ corresponds to an H\textsubscript{2}O atmosphere of $1.18 \times 10^{20}$ kg or a CO\textsubscript{2} atmosphere of $5.30 \times 10^{19}$ kg. By contrast, $\epsilon=1$ refers to planetary bodies with no atmosphere (i.e., a blackbody). The obtained cooling timescales are then interpolated for radii $0.1 \leq R \leq 1$ $R_{\oplus}$ and emissivities $0.01\leq \epsilon \leq 1$ using a 2-D surface fit.

Magma ocean lifetimes are also calculated for planets with a steam atmosphere \citep{Sol93}. For these models, we back-compute the effective emissivity as a function of time using the surface heat flux and surface temperature calculated by the interior model.  The emissivity decreases as a result of volatiles outgassing from the magma ocean into the atmosphere. For planetary radii 0.1 $R_{\oplus}$, 0.5 $R_{\oplus}$ and 1 $R_{\oplus}$, the mean emissivity is $\bar{\epsilon} \approx 0.001$, with a standard deviation of $\sigma \approx 10^{\mathrm{-4}}$. Since the average emissivities are comparable, we choose a fixed emissivity of $\epsilon = 0.001$ for the subsequent analyses of the steam atmosphere cases.

For each parameter combination, we estimate planetary brightness temperature for a given surface temperature and observation wavelength. Brightness temperature is a measure of the thermal radiation emitted by a planet and recorded by imaging instruments, and thus directly determines the potential for detection. Employing the definition of emissivity, i.e., the ratio of the spectral radiance at a given wavelength to that of a blackbody \citep[e.g.,][]{Catling&Kasting}, we map surface to brightness temperature using
\begin{linenomath*}
\begin{align} \label{eq:T_bright}
T_{\mathrm{b}}^{\mathrm{-1}}=\frac{k\lambda}{hc}\cdot \mathrm{ln}\left( \frac{e^{\mathrm{\frac{hc}{\lambda k T_{\mathrm{surf}}}}}-1}{\epsilon}+1 \right),
\end{align}
\end{linenomath*}
where $T_{\mathrm{b}}$ and $T_{\mathrm{surf}}$ are brightness and surface temperature, respectively, $k$ is Boltzmann's constant, $\lambda$ is wavelength, $h$ is Planck's constant, $c$ is the speed of light, and $\epsilon$ is the emissivity of the atmosphere.  The brightness temperature is then translated into an estimate of flux density for different time intervals after a giant impact.  We determine the time interval $\Delta t_{\mathrm{MO_{\mathrm{{\bar{R}}}}}}$ within which a magma ocean planet is bright enough to be detected (Eq. \ref{eq:master_equation_42}) by comparing the flux density evolution of a planet of radius $\bar{R}$ with a given telescope's sensitivity (Table \ref{tab:Perf}).

\section{Results and discussion}
\label{sec:results}

\subsection{Frequency of giant impacts during planet formation}
\label{sec:nbodyres}

\begin{figure}[tbh]
   \centering
   \includegraphics[width=0.49\textwidth]{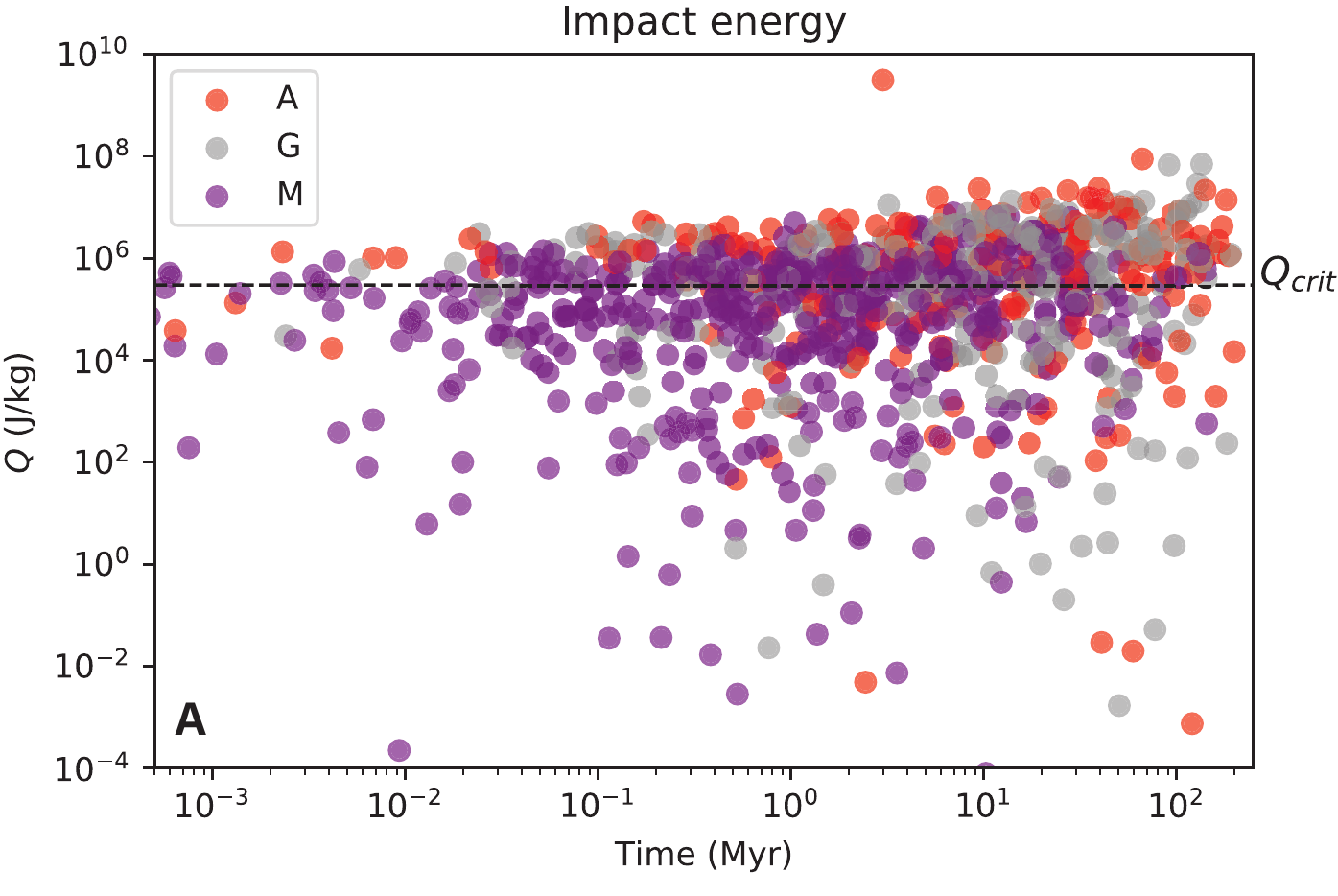}
   \includegraphics[width=0.45\textwidth]{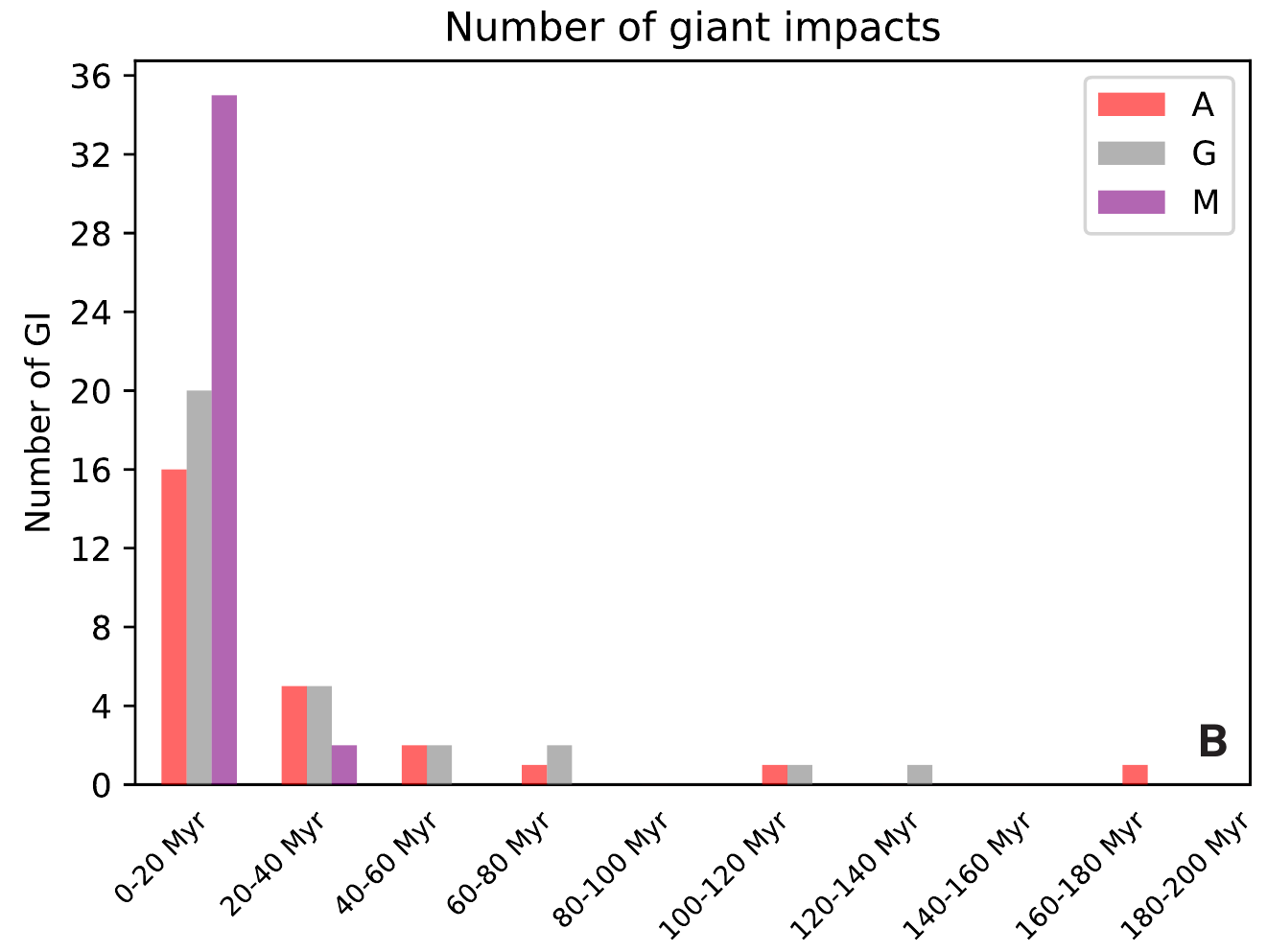}
   \captionsetup{justification=justified,font=small
	}
   \caption{\textbf{(A)} Specific impact energies for all simulations (A: A-stars, G: G-stars, M: M-stars) as a function of time. The dashed line denotes the energy threshold ($Q_{\mathrm{crit}} = 0.3 \times 10^{6}$ J/kg), above which we assume that a global magma ocean is generated from the impact. \textbf{(B)} The median number of giant impacts (GI) occurring within intervals of 20 Myr decays over time due to the decreasing number of planetesimals.}
   \label{fig:1}
\end{figure}

\begin{figure}[tbh]
   \centering
   \includegraphics[width=0.47\textwidth]{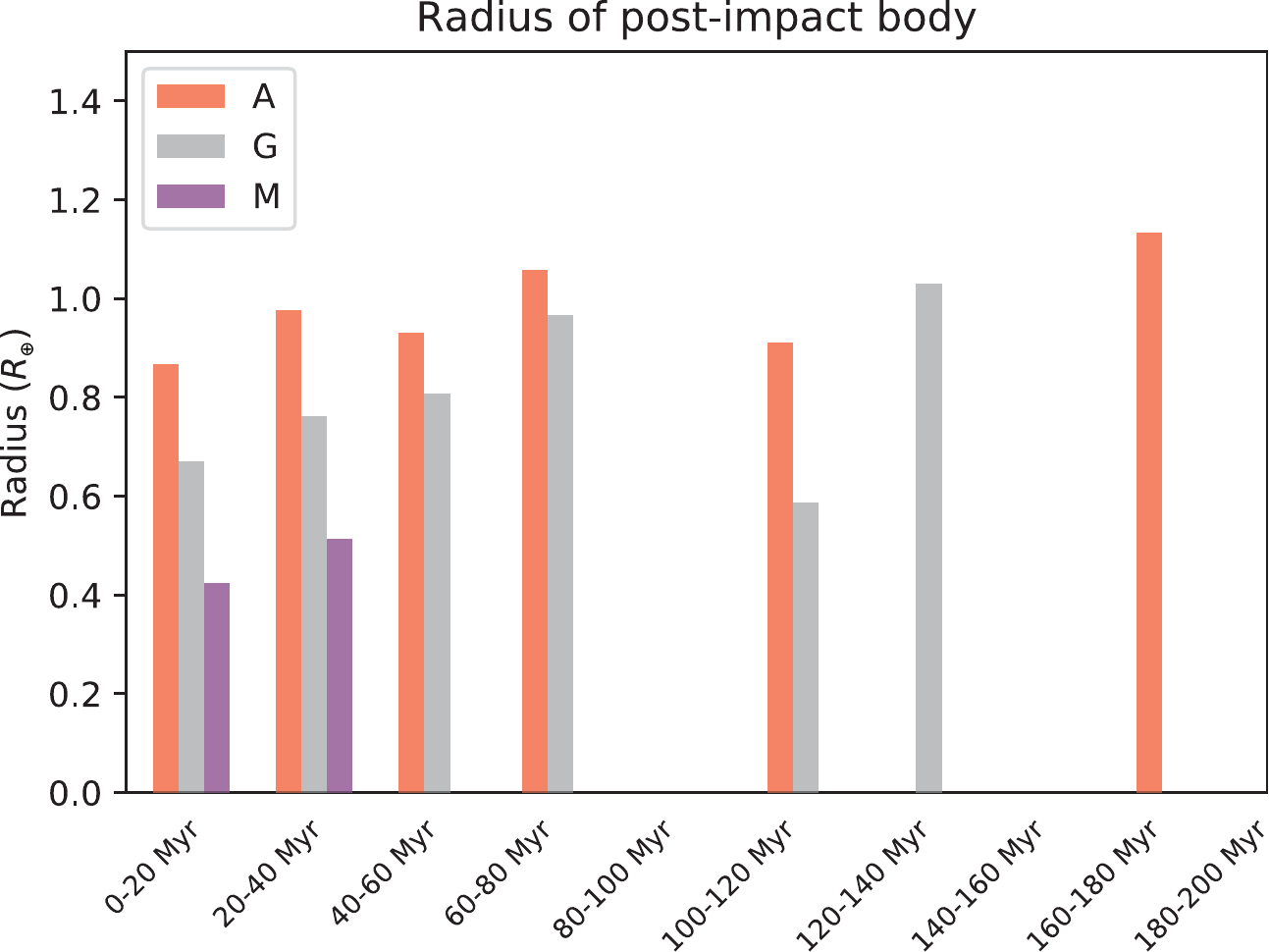}
   \captionsetup{justification=justified,font=small
	}
   \caption{Median radii of bodies generated from the giant impacts in Fig. \ref{fig:1}B for different time intervals of planet formation around different star types (A: A-stars, G: G-stars, M: M-stars). Planetary sizes are computed using a mass-radius relationship for rocky planets (Eq. \ref{eq:pythagoras3}).}
   \label{fig:12}
    \end{figure}

\begin{figure}[th]
	\centering
  \includegraphics[width=0.46\textwidth]{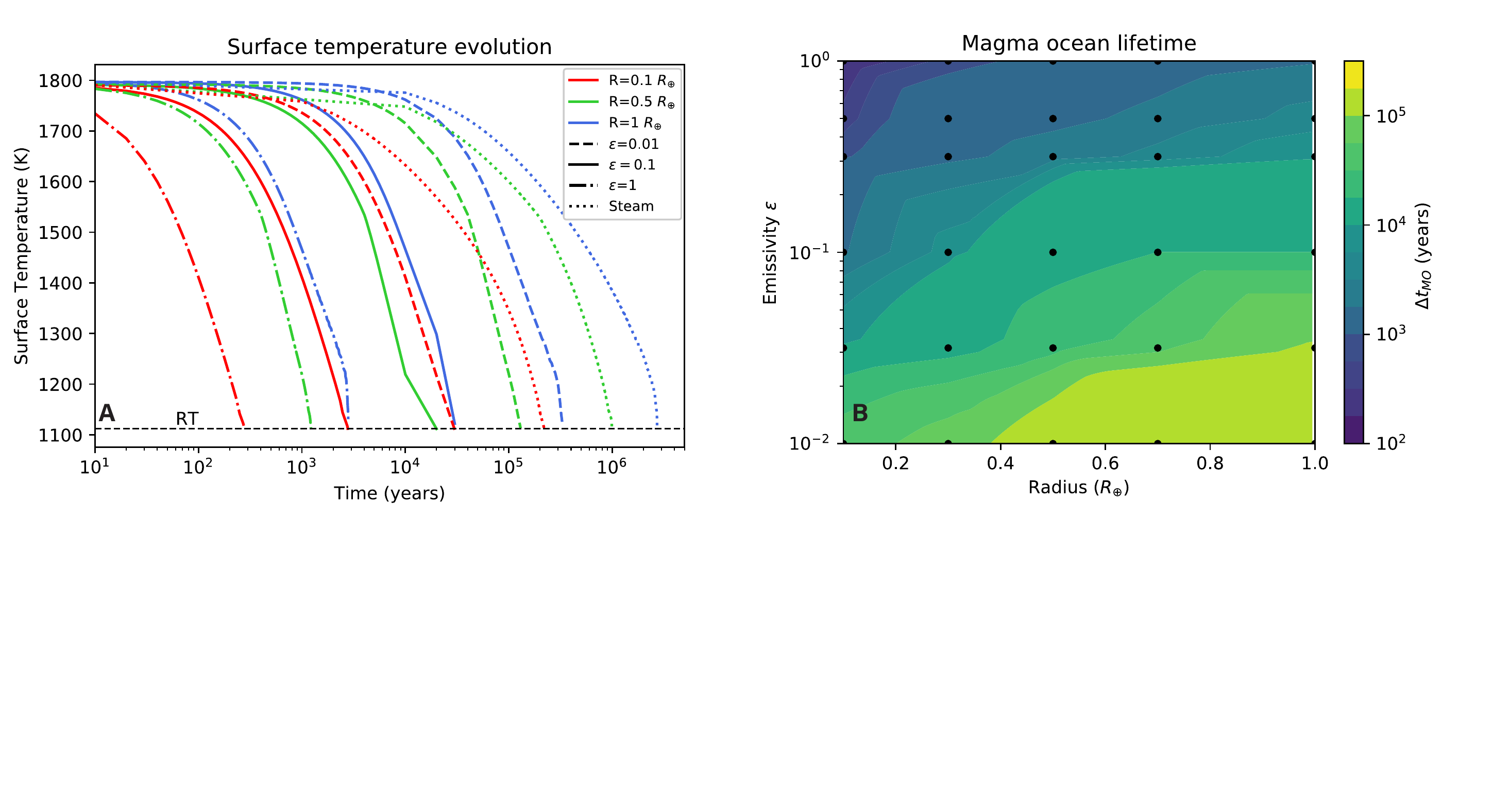}
  \includegraphics[width=0.48\textwidth]{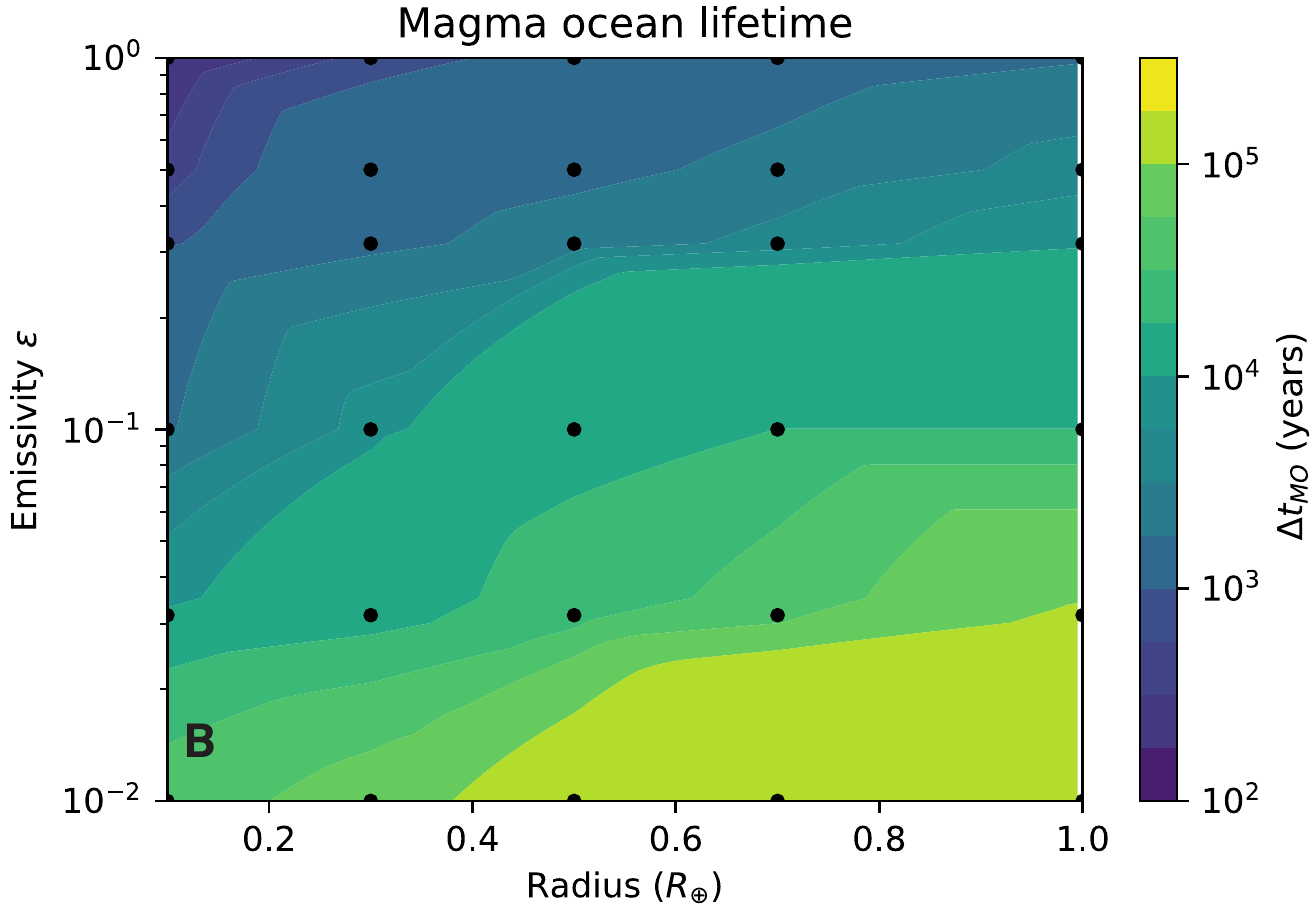}
	\captionsetup{justification=justified,font=small
	}
	\caption[Planetary surface temperature evolution and magma ocean lifetime]{\textbf{(A)} Surface temperature evolution of magma ocean planets with different radii and atmospheric emissivities (or steam atmosphere).  The horizontal dashed black line denotes the rheological transition (RT), below which a magma ocean may not be detectable due to a cold rigid lid forming at the surface.  Hence the RT criterion gives the minimum lifetime of a magma ocean that is likely to be detectable. \textbf{(B)} Magma ocean lifetimes $\Delta t_{\mathrm{MO}}$ obtained for the gray atmosphere model (black circles) and the 2-D surface fit for $0.1$ $R_{\oplus} \leq R \leq 1$ $R_{\oplus}$ and $0.01 \leq \epsilon \leq 1$.}
  \label{fig:2}
\end{figure}

For each spectral type (A, G, and M), we performed five \textit{N}-body simulations using the same initial conditions (Table \ref{tab:ic}). This allows us to constrain the statistical occurrence rate of magma ocean-inducing giant impacts as a function of time for a given star type. The specific impact energies of the collisions occurring in all simulations are shown in Fig. \ref{fig:1}A, together with the number of giant impacts for time intervals of 20 Myr shown in Fig.~\ref{fig:1}B. The latter are median values computed over all simulations for a given spectral type. The dashed horizontal line in Fig. \ref{fig:1}A marks the critical specific impact energy $Q_{\mathrm{{crit}}}= 0.3 \times 10^{6}$ J/kg. Most giant impacts take place within the first $\sim$20 Myr (Fig. \ref{fig:1}B), with similar trends for all stellar types. Simulations of systems forming around lower mass stars feature higher initial numbers of planetesimals and smaller protoplanetary disks, thus exhibiting an increased likelihood of close encounters and hence a greater number of giant impacts. We would like to point out that the time binning in 20 Myr intervals introduces a bias to stellar associations (like $\beta$ Pictoris with an age of 23 Myr) that lie close to edges of these bins. In addition, towards higher ages, the low number statistics for cases like AB Doradus (with an age of 150 Myr) affect the resulting number of giant impacts. Future work is needed to more realistically constrain the giant impact rate for the individual members of each association in order to overcome these limitations.

Previous work investigating the late stages of planet formation \citep[e.g.,][]{LeinhardtStewartI, Quintana} has shown that Earth-like bodies on average experience $\approx$10--20 giant impacts during accretion. The giant impact rates in our simulations are in good agreement with the aforementioned studies (especially for A- and G-stars), with the noted exception of the first 20 Myr bin. The early impact rates in $N$-body simulations are strongly dependent on the initial configuration of their constituent planetesimals. As it is notoriously difficult to construct dynamically self-consistent initial architectures for $N$-body systems, each simulation will experience an initial period of dynamical instability. The high rate of impacts in the first 20 Myr bin reflects this initial period of instability and should therefore be taken as an overestimate. However, while the first 20 Myr bin is biased towards higher impact rates, in general our study tends to underestimate the giant impact rate due to the way collisions are handled. By assuming perfect merging, we are likely underestimating the giant impact rate \citep{Quintana}. 

Our simulations and those performed in \citet{LeinhardtStewartI} and \citet{Quintana} begin from different sets of initial conditions, including different numbers and masses of planetesimals. Thus, the dynamical configurations of the systems are not identical and the initial stages of the simulations are expected to differ, which in particular influences the early impact rate. This can also be interpreted as starting from different stages in the planet formation process. As computing power increases, the early giant impact rate following disk dispersal will become better constrained by studies starting from an increasing number of smaller bodies, allowing the simulated systems to reach more realistic dynamical configurations prior to the onset of the giant impact phase. Future studies will also benefit from the ability to run an increasing number of simulations for each initial condition, allowing researchers to better understand the effect of stochasticity on the giant impact rate.

The radii of the post-impact bodies are shown in Fig. \ref{fig:12}. In general, perfect accretion of the colliding bodies results in planets with masses of less than $1.5~M_{\oplus}$ and radii less than $1.1~R_{\oplus}$. Protoplanetary disk mass scales with stellar mass and, therefore, the amount of material available for planet formation is greater around higher-mass stars (Table \ref{tab:ic}). The post-impact bodies around A-type stars are thus expected to be larger and more massive on average. In terms of detectability, larger planets are favored as they are likely to emit an increased amount of thermal radiation and have extended magma ocean lifetimes.
The specific impact energies are normalized by the total mass of the colliding bodies. Hence, even though post-impact bodies are on average smaller around lower-mass stars, they still experience similar specific impact energies to planets around A-type stars (Fig. \ref{fig:1}A).

\subsection{Magma ocean cooling timescales}

Magma ocean lifetimes are important to assess the direct detectability of planets in their early evolution stage, and determine which young stellar associations can be considered for future direct observations. As an example, if the magma ocean lifetime is on the order of millions of years or shorter, young stellar associations with ages $< 100$ Myr are the prime target for observations.  The evolution of surface temperature and resulting magma ocean lifetimes are shown in Fig. \ref{fig:2}A and Fig. \ref{fig:2}B, respectively. We track the lifetime of the magma ocean from an initially fully molten state until the time when the rheological transition (40\% melt fraction) reaches the surface. Larger planets experience longer magma ocean lifetimes due to their reduced surface area to volume ratio (i.e., a larger energy budget relative to radiative heat loss). Moreover, the emissivity strongly influences magma ocean cooling timescales. A low emissivity is indicative of a thick outgassed atmosphere, whereas a high value is indicative of a thin atmosphere. For a given planet size, a decreasing emissivity from $\epsilon=1$ to $\epsilon=0.01$ causes the cooling timescale to increase by $\sim$2 orders of magnitude. The longest magma ocean lifetimes are observed for cases with a steam atmosphere ($\approx$ 2--3 $\times$ $10^{6}$ years).

Previous studies of coupled interior-atmosphere evolution reveal typical magma ocean lifetimes of $10^{\mathrm{-2}}-10^{\mathrm{6}}$ years \citep{ET08}, and possibly up to hundreds of millions of years \citep{Hamanonat,Hamano}. Our magma ocean lifetimes ($10^{\mathrm{2}}-3 \times 10^{\mathrm{6}}$ years) compare well to these previous studies and are typically slightly shorter due to the different definition of the cooling timescale.  \citet{ET08} and \citet{Hamanonat} define the magma ocean lifetime based on the time required for a fully molten magma ocean to reach the solidus. In our modeling, we calculate the time taken for the rheological transition to reach the surface. Beyond this time, the luminosity of the planet is uncertain since it is unclear whether the planetary surface remains hot and molten or if a cold rigid lid forms. Hence, by design, we compute conservative estimates of magma ocean detectability based on minimum estimates of magma ocean lifetimes.  This is further compatible with our exclusion of internal heat sources, which would extend magma ocean lifetimes.

\subsection{Telescope performance}
\label{sec:perf}

\begin{figure*}[th]
  \centering
  \includegraphics[width=\textwidth]{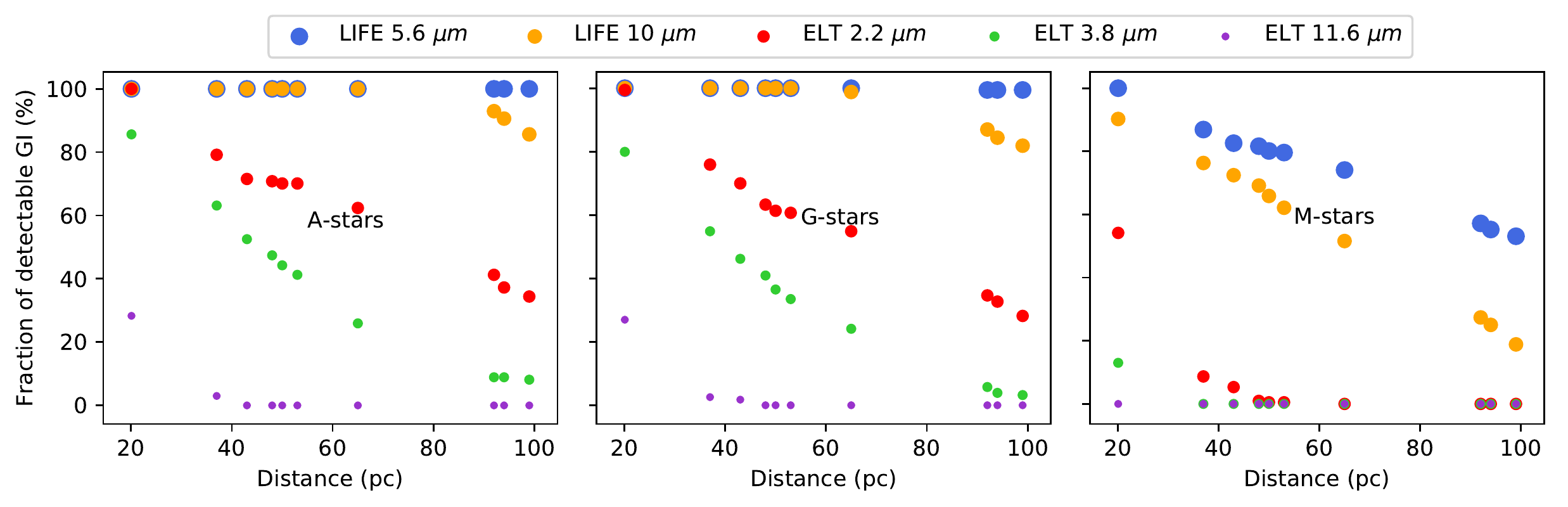}
  \captionsetup{justification=justified,font=small
  }
  \caption[FGI]{Fraction of giant impacts (GI) occurring around A-, G-, and M-stars that can be detected by LIFE and ELT. Here, we only consider the IWA of the instruments without taking into account the effect of the planetary atmosphere and the exposure time. The positions of the colored data points on the x-axis correspond to the distances of the young stellar associations in Table \ref{tab:SA}. The orbital range that can be resolved by a telescope decreases with increasing distance from the Sun. Hence, the fraction of giant impacts that can be detected is smaller for stellar associations located farther away. Due to its smaller IWA, and thus higher angular resolution,  LIFE can  image higher fractions of impacts.}
  \label{fig:3}
\end{figure*}
\begin{figure*}[p!]
  \centering
  \includegraphics[width=0.7\textwidth]{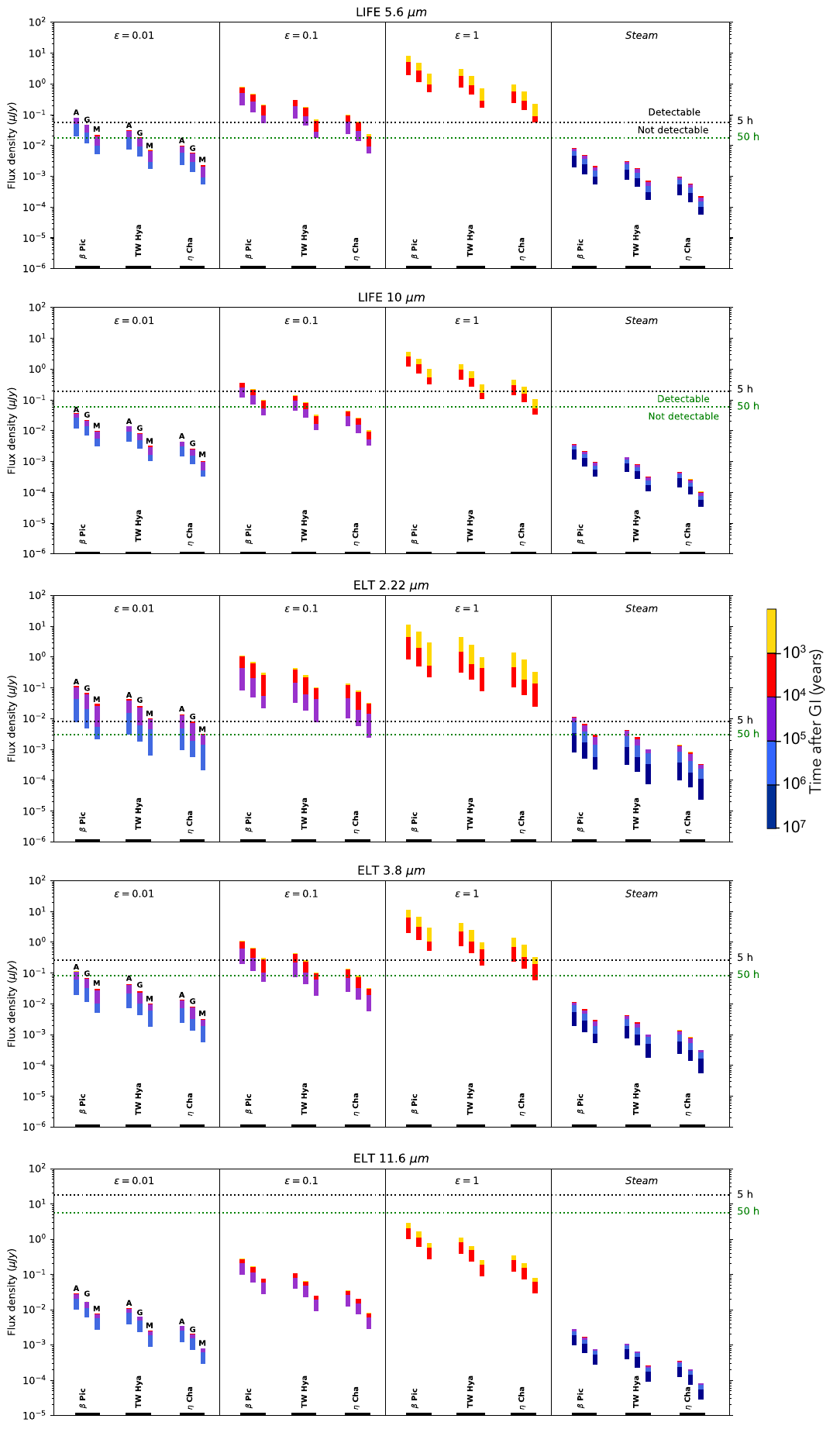}
  \captionsetup{justification=justified,font=small
  }
\caption[Sens]{Flux density evolution for planets located within the $\beta$ Pictoris (37 pc), TW Hydrae (53 pc), and $\eta$ Chamaeleontis (94 pc) associations during the cooling of a magma ocean. The subplots and subplot panels indicate different telescope filter wavelengths and atmospheric emissivities, respectively. Each set of bars consists of flux evolutions for bodies located inside a given stellar association, orbiting central stars of different spectral type (A, G or M). The colors indicate different time intervals after the occurrence of a giant impact (GI).  Each vertical bar terminates according to the cooling timescale of the magma ocean $\Delta t_\mathrm{MO}$ (see Fig. \ref{fig:2}A and Fig. \ref{fig:2}B). Planetary sizes used for the flux calculations are taken from the data in Fig. \ref{fig:12}, based on the spectral type of the host star and the age of the considered stellar association. The black and green horizontal dashed lines mark the sensitivity of the telescope filters for integration times of 5 and 50 hours (Table \ref{tab:Perf}), respectively. Magma ocean bodies displaying fluxes greater than these thresholds are detectable.}
  \label{fig:4}
\end{figure*}

The IWA and sensitivity of the ELT filters and LIFE (Table \ref{tab:Perf}) are taken as detectability thresholds for constraining the observability of giant collision afterglows. Using the results of the $\emph{N}$-body simulations (Sect.~\ref{sec:nbodyres}), we determine the fraction of detectable giant impacts within the young stellar groups (Table~\ref{tab:SA}). These have ages ranging from 10 to 150 Myr and are thus associated with different numbers of collisions and post-impact body characteristics according to the time bins in Fig. \ref{fig:1}B and Fig. \ref{fig:12}. Figure \ref{fig:3} illustrates the fraction of detectable giant impacts around A-, G-, and M-type stars as a function of distance from the Sun. We only consider the IWA of the instruments, thus not accounting for atmospheric effects and exposure times, as well as the ages of the stellar associations.  For each telescope filter, a data point at a given distance corresponds to one of the ten stellar associations in Table \ref{tab:SA}.

Due to its small IWA and hence high angular resolution, LIFE is able to capture giant impacts that are close ($\approx 0.1$ AU) to the central star. Therefore, compared to the ELT, it displays a shallow decrease in the amount of detectable impacts with increasing distance from the Sun. Furthermore, the baseline of the individual free-flying LIFE telescopes could be increased, thus achieving higher spatial resolution. This would enable the imaging of the innermost region of planet forming regions. A- and G-stars both exhibit a similar decay in the number of observable giant impacts. Conversely, for M-type stars the fraction of detectable collisions decreases more rapidly due to the smaller radial extent of the disk (Table \ref{tab:ic}). This causes giant collisions to occur close to the central star, rendering the task of imaging them challenging for LIFE and almost impossible for the ELT, especially at longer wavelengths. 

During magma ocean cooling, the outgoing radiation flux decays due to the decrease of surface temperature. Figure \ref{fig:4} shows flux evolutions following a giant impact for magma ocean bodies located inside the stellar associations $\beta$ Pictoris (37 pc), TW Hydrae (53 pc), and $\eta$ Chamaeleontis (94 pc), as they reasonably sample the distance range (up to 100 pc) considered here. Planetary sizes are extracted from Fig. \ref{fig:12}, according to the spectral type of the host star and the age of the stellar association. The energy radiated by a planet is quantified by its brightness temperature (Eq. \ref{eq:T_bright}), which depends on the atmospheric emissivity; surface and brightness temperature are equivalent for a planet without an atmosphere. For observation times of five hours and emissivities $\epsilon= 1$, planetary bodies within the three associations are detectable over the entirety or majority of the lifetime of a magma ocean, except for ELT 11.6 $\mu m$. However, these planets tend to cool faster, thus displaying shorter time windows during which they are observable ($\Delta t_{\mathrm{MO_{\mathrm{\bar{R}}}}}$, see Eq. \ref{eq:master_equation_42}). By contrast, bodies with lower atmospheric emissivities or a steam atmosphere never or barely exceed the telescopes' sensitivity thresholds, but could be observable for an extended duration.

If the observation time is increased by a factor of ten (50 hours instead of 5 hours), the sensitivity thresholds decrease by a factor $\sqrt{10}$, leading to an improved detection of magma oceans. For observations of 50 hours, most planets located within TW Hydrae, and $\eta$ Chamaeleontis would still not be detectable due to their distance from the Sun and hence apparent faintness. However, most of the telescope filters would be able to observe magma oceans within $\beta$ Pictoris at least once during their lifetimes. Magma ocean bodies with or without an atmosphere are too faint to be detected with the ELT at 11.6 $\mu$m due to high background emission at these wavelengths. 

\subsection{Likelihood of detecting magma ocean planets}
\label{sec:likelihood}

\begin{figure*}[th]
      \centering
      \includegraphics[width=\textwidth]{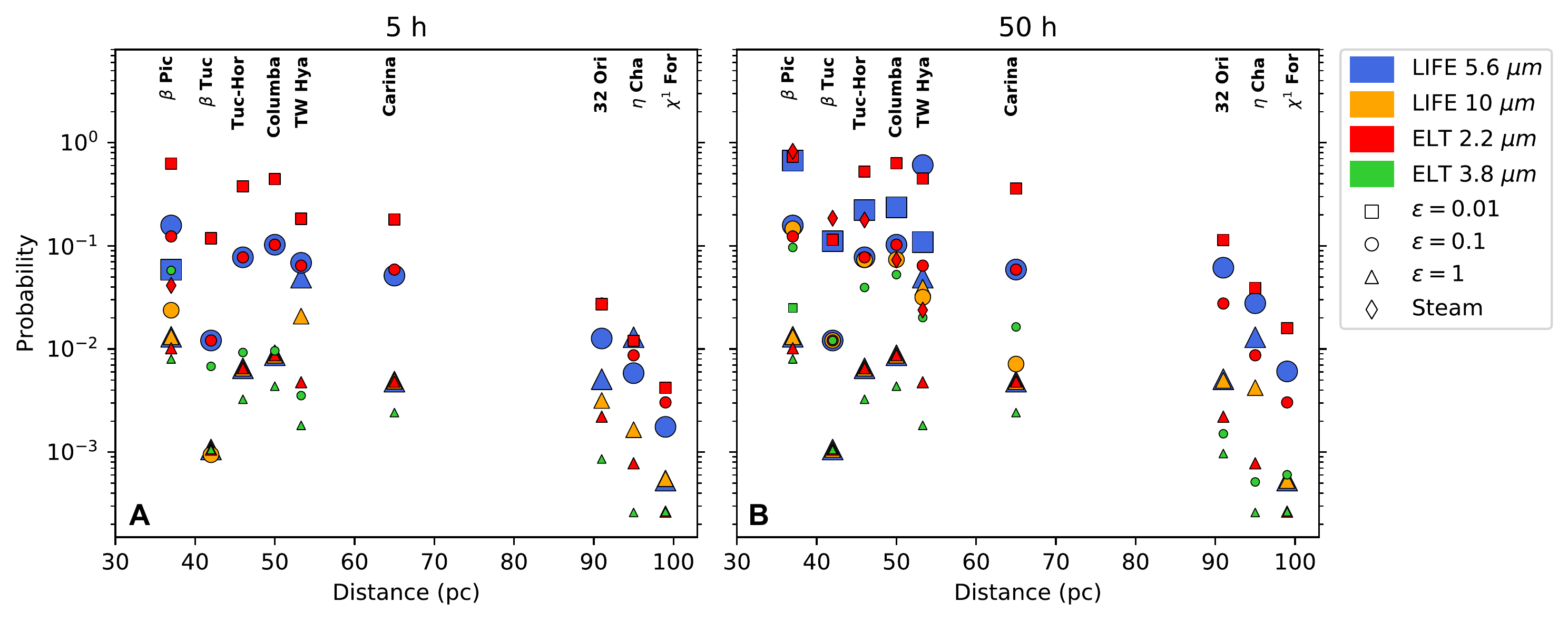}
      \captionsetup{justification=justified,
font=small}
      \caption[Probability of detecting a magma ocean planet in each stellar association]{Probability of detecting at least one magma ocean planet in nearby young stellar associations (Table \ref{tab:SA}) for observation times of \textbf{(A)} 5 hours and \textbf{(B)} 50 hours. The colors and shapes refer to telescope filters and atmospheric emissivities, respectively. The prime stellar target for the detection of magma oceans is the $\beta$ Pictoris association (37 pc, 23 Myr), followed by Columba (50 pc, 42 Myr), TW Hydrae (53 pc, 10 Myr), and Tucana-Horologium (48 pc, 45 Myr). Due to the low probability of giant impacts at their respective ages (Fig. \ref{fig:1}B), no data points are shown for stars located inside the AB Doradus association (20 pc, 150 Myr). ELT 11.6 $\mu m$ is not able to detect any magma ocean planets.}
      \label{fig:5}
\end{figure*}

\setlength{\tabcolsep}{9pt} 
\begin{table*}[bh!]
	\centering
	\begin{tabular}{c| c c c c c c c c c c} 
		\toprule
		\toprule
		\multicolumn{5}{r}{\textbf{\hspace{3cm}Probability (\%) - 5 h}} & \multicolumn{5}{r}{\textbf{\hspace{3cm}Probability (\%) - 50 h}}\\
		\cmidrule{3-6}
		\cmidrule{8-11}
		\textbf{Filter}&\textbf{SA}&$\epsilon=0.01$&$\epsilon=0.1$&$\epsilon=1$&Steam& &$\epsilon=0.01$&$\epsilon=0.1$&$\epsilon=1$&Steam \\
		\midrule
		\parbox[t]{6mm}{\multirow{4}{*}{{L56}}}
		&$\beta$ Pictoris&5.9&15.8&1.3&-&&66.9&15.8&1.3&-\\
		&Tucana-Horologium&-&7.7&0.6&-&&22.2&7.7&0.6&-\\
		&Columba&-&10.2&0.9&-&&23.6&10.2&0.9&-\\
		&TW Hydrae&-&6.8&4.9&-&&10.9&60.8&4.9&-\\
		\midrule
		\parbox[t]{6mm}{\multirow{4}{*}{{L10}}}
		& $\beta$ Pictoris&-&2.4&1.3&-&&-&14.6&1.3&-\\
		&Tucana-Horologium&-&-&0.6&-&&-&7.4&0.6&-\\
		&Columba&-&-&0.9&-&&-&7.4&0.9&-\\ 
		&TW Hydrae&-&-&2.1&-&&-&3.2&4.0&-\\
		\midrule
		\parbox[t]{6mm}{\multirow{4}{*}{{E22}}}
		& $\beta$ Pictoris&62.8&12.3&1.0&4.1&&73.1&12.3&1.0&82.6\\
		&Tucana-Horologium&37.8&7.7&0.6&-&&52.5&7.7&0.6&17.8\\
		&Columba&44.4&10.2&0.9&-&&63.1&10.2&0.9&7.3\\
		&TW Hydrae&18.3&6.4&0.5&-&&44.9&6.4&0.5&2.4\\
		\midrule
		\parbox[t]{6mm}{\multirow{4}{*}{{E38}}}
		& $\beta$ Pictoris&-&5.8&0.8&-&&2.5&9.6&0.8&-\\
		&Tucana-Horologium&-&0.9&0.3&-&&-&3.9&0.3&-\\
		&Columba&-&1.0&0.4&-&&-&5.3&0.4&-\\
		&TW Hydrae&-&0.3&0.2&-&&-&2.0&0.2&-\\
		\bottomrule
	\end{tabular}
	\captionsetup{justification=justified,font=small
	}
	\caption{Probability of detecting at least one magma ocean planet in the most promising stellar associations ($\beta$ Pictoris, Tucana-Horologium, Columba, and TW Hydrae) for different integration times (5 h and 50 h per star), instruments (L56: LIFE 5.6 $\mu m$, L10: LIFE 10 $\mu m$, E22: ELT 2.2 $\mu m$, E38: ELT 3.8 $\mu m$), and atmospheric emissivities (or steam atmosphere). We do not show detection probabilities for ELT 11.6 $\mu m$, as it is not able to detect any magma ocean planets (cf. Fig. \ref{fig:4}).}
	\label{tab:nolupu}
\end{table*}

\setlength{\tabcolsep}{6pt}   
\begin{table}[bh!]
	\centering
	\begin{tabular}{c| c c c} 
		\toprule
		\toprule
		\textbf{SA}&\textbf{Star (Spec. type)}&K-mag.&Flux density (Jy)\\
		\midrule
 		\parbox[t]{9mm}{\multirow{3}{*}{{$\beta$ Pic}}}
 		&HR 6070 (A1)&4.74&8.52\\
 		&HIP 10679 (G2)&6.26&2.10\\
        &HIP 11437B (M0)&7.92&0.46\\
		\bottomrule
	\end{tabular}
	\captionsetup{justification=justified,font=small
	}
	\caption{K band brightnesses of $\beta$ Pictoris moving group A-, G-, and M-type stars located close to the assumed mean distance \citep{Beta}. K band magnitudes are from the SIMBAD database \citep{Simbad}.}
	\label{tab:magn}
\end{table}

The probabilities of detecting at least one magma ocean planet in young nearby stellar associations (Table \ref{tab:SA}) are shown in Fig. \ref{fig:5} for integration times of 5 hours and 50 hours, for various combinations of atmospheric emissivities and instruments. These were calculated according to Eq. \ref{eq:master_equation_42}. A young stellar age translates to a high number of expected giant impacts (Fig. \ref{fig:1}B). Therefore, the younger a stellar association and the more stars of a given type it contains, the higher the probability of detecting at least one magma ocean planet. Bodies with lower emissivities are the most likely to be directly observed due to their long-lived magma oceans, but the atmosphere could be sufficiently dense to prevent the near surface from being imaged. In contrast, a less dense atmosphere will pose a lower barrier to observing the surface, but it will enable the magma ocean to cool faster.

Generally, the likelihoods of detecting hot molten planets within observation times of five hours (Fig. \ref{fig:5}A) are highest for the $\beta$ Pictoris association, due to its proximity to the Sun, its relatively high number of stellar members and its young age (Table \ref{tab:SA}). Beyond this stellar group, mostly bodies with high atmospheric emissivities (i.e., higher brightness temperatures) can be detected. If planetary fluxes for a given atmospheric emissivity are higher than a given instrument's sensitivity over the whole magma ocean cooling timescale (e.g., for LIFE 5.6 $\mu m$, $\epsilon=1$), the highest probabilities are obtained for the association containing more stars (Table \ref{tab:SA}) and/or experiencing the highest number of detectable giant impacts  (Fig. \ref{fig:3}). The preferred instruments for the detection of magma oceans are ELT 2.2 $\mu m$, followed by LIFE 5.6 $\mu m$ (see Table~\ref{tab:nolupu}). This assumes that the full number of young planetary systems in star-forming stellar regions can be utilized.  If longer integration times of 50 hours (Fig. \ref{fig:5}B) are allocated, the prospects of detection are improved, especially for planets with low atmospheric emissivities. In both integration scenarios, the most promising stellar targets to be explored are $\beta$ Pictoris (37 pc, 23 Myr), Columba (50 pc, 42 Myr), TW Hydrae (53 pc, 10 Myr), and Tucana-Horologium (48 pc, 45 Myr).

ELT 2.2 $\mu m$ is the preferred filter for future magma ocean observations. We remind the reader, however, that in our assumptions for ELT K band observations we assumed that the instrument is equipped with a laser tomography adaptive optics (LTAO) system (see Table~\ref{tab:Perf}). Changing this assumption to seeing limited performance or ground-layer adaptive optics (AO) significantly reduces the achieved sensitivity. In addition, at such short wavelengths, it is important to compare the contrast between the stars in a given association and the magma ocean planets. We therefore convert the K band magnitudes of three stars (one A-, one G-, and one M-type star) from the $\beta$ Pictoris moving group, the most promising stellar group for magma ocean detections, into flux densities (Table \ref{tab:magn}). We find that stars are around 5-6 orders of magnitude brighter than magma ocean bodies shortly after the occurrence of a giant impact in the best case scenario (i.e., for $\epsilon=1.$) and the contrast further increases by factors of ten going to $\epsilon=0.1$, $\epsilon=0.01$ and a steam atmosphere. While it is not trivial to achieve this performance from the ground at the ELT at 2.2 $\mu$m, dedicated high-contrast instruments equipped with extreme AO systems aim at achieving near-infrared contrast levels of 10$^{-8}$-10$^{-9}$ \citep[e.g.,][]{kasper2010}.

Finally, it is worth pointing out that while we were interested in the detectability of magma oceans in a statistical sense, there are individual stars in each stellar association that have an intrinsically higher probability than others given their smaller-than-average distance. This is particularly true for $\beta$ Pictoris moving group members that show a large dispersion in their distances. 

\subsection{Comparison with atmospheric calculations}

\begin{figure*}[th]
      \centering
      \includegraphics[width=\textwidth]{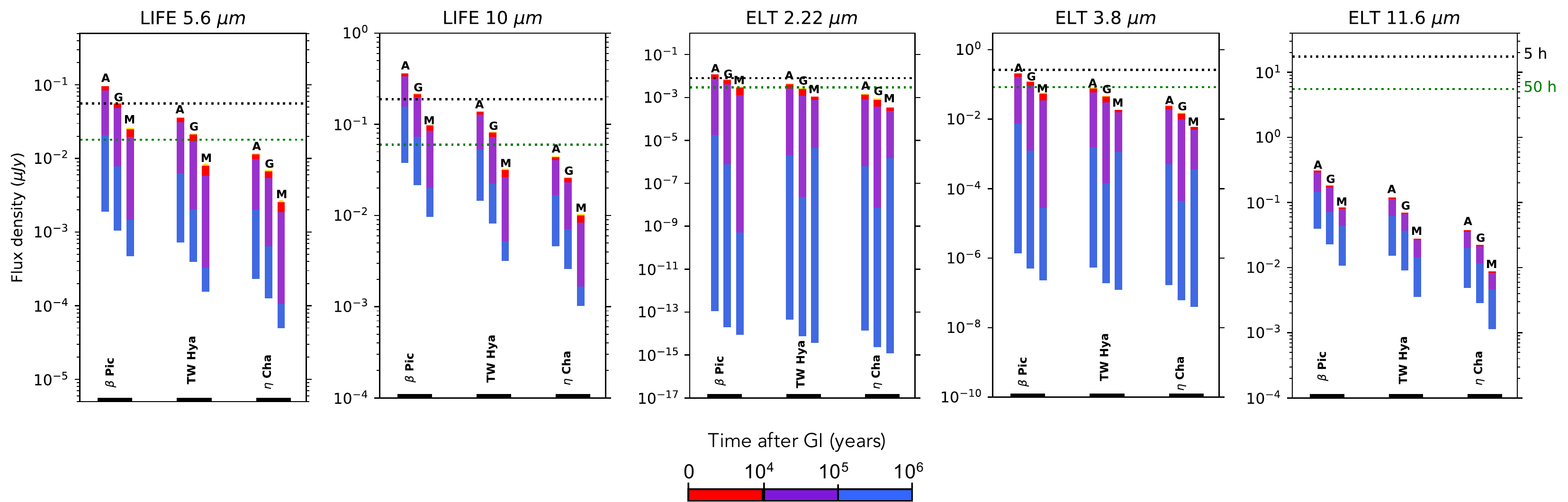}
      \captionsetup{justification=justified,
font=small}
      \caption[Flux densities using interpolation from Lupu data]{Flux density evolutions during the cooling of a magma ocean with an atmospheric emissivity $\epsilon=0.01$, computed using brightness temperatures from \citet[][Fig. 7]{Lupu}. As for Fig. \ref{fig:4}, planets in the $\beta$ Pictoris (37 pc), TW Hydrae (53 pc), and $\eta$ Chamaeleontis (94 pc) associations are considered. The subplots indicate different telescope filter wavelengths. Each set of bars consists of flux evolutions for bodies in a given stellar association, orbiting central stars of spectral type (A, G or M). The colors indicate different time intervals after the occurrence of a giant impact (GI). Each vertical bar terminates according to the cooling timescale of the magma ocean $\Delta t_\mathrm{MO}$ (see Fig. \ref{fig:2}B). Planetary sizes used for the flux calculations are taken from the data in Fig. \ref{fig:12}, based on the spectral type of the host star and the age of the considered stellar association. The black and green horizontal dashed lines mark the sensitivity of the telescope filters for integration times of 5 and 50 hours (Table \ref{tab:Perf}), respectively. Magma ocean bodies displaying fluxes greater than these thresholds are detectable.}
      \label{fig:8}
\end{figure*}

\setlength{\tabcolsep}{6.3pt}
\begin{table}[th!]
	\centering
	\begin{tabular}{c| c c c} 
		\toprule
		\toprule
		&&{\textbf{\underline{P (\%) - 5 h}}} & {\textbf{\underline{P (\%) - 50 h}}}\\
		\textbf{Filter}&\textbf{SA}&$\epsilon=0.01$&$\epsilon=0.01$\\
		\midrule
 		\parbox[t]{6mm}{\multirow{4}{*}{{L56}}}
 		&$\beta$ Pictoris&4.4&35.7\\
		&Tucana-Horologium&-&14.0\\
        &Columba&-&16.5\\
        &TW Hydrae&-&7.1\\        
        \midrule
        \parbox[t]{6mm}{\multirow{4}{*}{{L10}}}
 		&$\beta$ Pictoris&26.4&86.1\\
        &Tucana-Horologium&-&25.3\\
        &Columba&-&28.6\\
        &TW Hydrae&-&13.1\\
		\midrule
        \parbox[t]{6mm}{\multirow{4}{*}{{E22}}}
 		&$\beta$ Pictoris&0.1&7.1\\
        &Tucana-Horologium&-&0.4\\
        &Columba&-&0.4\\
        &TW Hydrae&-&0.04\\
		\midrule
        \parbox[t]{6mm}{\multirow{2}{*}{{E38}}}
 		&$\beta$ Pictoris&-&5.1\\
        &Tucana-Horologium&-&0.06\\
		\bottomrule
\end{tabular}
	\captionsetup{justification=justified,font=small
	}
	\caption{Probability of detecting at least one magma ocean planet with atmospheric emissivity $\epsilon=0.01$, based on flux density calculations performed with brightness temperatures from \citet{Lupu} (see Fig. \ref{fig:8}). Detection likelihoods are shown for the most promising stellar targets, for different integration times (5 h and 50 h per star) and instruments (L56: LIFE 5.6 $\mu m$, L10: LIFE 10 $\mu m$, E22: ELT 2.2 $\mu m$, E38: ELT 3.8 $\mu m$). We do not show detection probabilities for ELT 11.6 $\mu m$, as it is not able to detect any magma ocean planets (cf. Fig. \ref{fig:8}).}
	\label{tab:5}
\end{table}

We compare flux density evolutions and detection probabilities presented in Sect. \ref{sec:perf} and Sect. \ref{sec:likelihood} with those obtained using data based on radiative and convective equilibrium (atmospheric) calculations. Figure 7 in \citet{Lupu} shows brightness temperature as a function of wavelength, for surface temperatures of 1200 K, 1600 K, and 2200 K and surface pressures of $p_{\mathrm{surf}}=10$ bar and $p_{\mathrm{surf}}=100$ bar. These relations are derived by considering an atmosphere in equilibrium with a magma ocean that has a composition similar to bulk silicate Earth (BSE). Using Eq. 9 in \citet{ET08}, we calculate the mass of an H\textsubscript{2}O atmosphere and its associated surface pressure for atmospheric emissivities of $\epsilon=0.001$, $\epsilon=0.01$ and $\epsilon=0.1$.  We consider an H\textsubscript{2}O atmosphere since this is the dominant molecule in the \citet{Lupu} model.

For $\epsilon=0.001$ and $\epsilon=0.1$, we obtain surface pressures outside the range considered by \citet{Lupu} (i.e., $p_{\mathrm{surf}}=232$ bar and $p_{\mathrm{surf}}=2$ bar, respectively).  However, for $\epsilon=0.01$ we compute $p_{\mathrm{surf}}=23$ bar, and by linearly interpolating the results of \citet{Lupu} we map surface to brightness temperature.  Flux density evolutions for this alternative model are presented in Fig. \ref{fig:8} for ELT and LIFE filters, and the detection probabilities are summarized in Table \ref{tab:5} for the most promising stellar targets. While ELT is the preferred instrument in the previous scenario (using Eq. \ref{eq:T_bright} for the calculation of brightness temperature), in this alternative model the best observation strategy is represented by both LIFE 10 $\mu m$ and LIFE 5.6 $\mu m$. The ELT filters yield substantially lower chances for detection. This dependence on the chosen relation between surface temperature and brightness temperature provides additional motivation to develop fully-coupled models of the interior, surface, and atmosphere to track the evolution of a rocky body during the magma ocean stage. 

\section{Summary and conclusion}
\label{sec:conclusions}

During the final stages of terrestrial planet formation, growing planets experience violent pair-wise collisions, known as giant impacts. These collisions release a substantial amount of energy, which can lead to extensive melting of planetary bodies. The post-impact surface magma ocean may be detectable as an \quotes{afterglow}. We present a modeling framework that provides first-order predictions of the detectability of such events using performance estimates of next-generation direct imaging instruments, simulations of giant impact occurrence rates during planet formation around A-, G-, and M-stars, and models of magma ocean cooling for planets of various sizes and atmospheric emissivities.  With sufficient observing time to fully exploit the number of planetary systems in nearby stellar associations, the observability of giant impacts dominantly depends on two parameters: (1) the expected number of giant impacts for a given stellar age, and (2) the magma ocean lifetime. The latter crucially depends on the atmospheric emissivity and hence the mass and composition of the planetary atmosphere, which controls the rate at which a magma ocean can lose its heat to space. Planets with high emissivities (i.e., thinner atmospheres) are easier to detect, but at the expense of a shorter magma ocean lifetime. Conversely, dense atmospheres delay the cooling of magma oceans but will also hinder the imaging of planetary surfaces. 

We find that target selection favoring young and nearby stellar associations, which contain a large number of stars, significantly increases the likelihood of detecting a magma ocean planet. Furthermore, for sufficiently long integration times, the $\beta$ Pictoris association is best suited for potential future observations of molten protoplanets, followed by Columba, TW Hydrae, and Tucana-Horologium.  Within the modeling framework employed in this work, K band filters of the ELT and mid-infrared wavelength filters of a space-based interferometer similar to LIFE are  best suited for the exploration of protoplanetary collision afterglows. 

Our work highlights the need to explore observing strategies specifically devised for the stars (and their associations) that are most likely to host molten planetary bodies. For integration times of 50 hours, we predict that detection probabilities can reach values as high as $\gtrsim 60\%$ for young stellar groups, such as $\beta$ Pictoris, depending on the statistical distribution of atmospheric properties and volatile constituents of molten protoplanets. Our estimates can be further enhanced by constraining the expected distribution of planetary (and thus atmospheric) compositions within the suggested stellar associations, and by providing more precise stellar ages and estimates of their total mass. Finally, an optimal observation strategy can be devised by focusing on the most prospective stellar members of the considered associations, and further constraining the anticipated giant impact rate per individual star in a stellar group.

Our work motivates the use of ELT instruments to search for molten rocky protoplanets, and highlights the advantages of a space-based mid-infrared nulling interferometer similar to LIFE for future explorations devoted to the study of the formation and early evolution of planetary bodies. Such a telescope would complement upcoming missions that use both direct and indirect exploration approaches for exoplanet science.

\begin{acknowledgements}
The authors thank the referee Markus Kasper for a timely and constructive report that lead to considerable improvements of the manuscript. We would like to thank Taras Gerya for helpful discussions and for reading an earlier version of the manuscript. IB thanks Matthieu Laneuville for valuable suggestions, and acknowledges financial support from a scholarship by the Japanese Ministry of Education, Culture, Sports, Science and Technology (MEXT). IB and TL thank Patrick Sanan for technical guidance on using the \textsc{SPIDER} code. TL thanks Michael Meyer for his persistent enthusiasm about protoplanet collision afterglows and acknowledges financial support from ETH Zurich Research Grant ETH-17 13-1, a MERAC travel grant of the Swiss Society for Astrophysics and Astronomy, and a postdoctoral fellowship of the Swiss National Science Foundation (SNSF). DJB acknowledges SNSF Ambizione Grant 173992. Parts of this work have been carried out within the framework of the National Center for Competence in Research PlanetS supported by the SNSF. This work made use of infrastructure provided by S3IT (\href{www.s3it.uzh.chuzh}{www.s3it.uzh.ch}), the Service and Support for Science IT team at the University of Zurich. The numerical simulations were analyzed using the open source software environment \textsc{matplotlib} \citep[][\href{https://matplotlib.org}{matplotlib.org}]{matplotlib} and the \textsc{astropy} package \citep{astropy18}.
\end{acknowledgements}

\bibliographystyle{aa} 

\begin{thebibliography}{65}
	\expandafter\ifx\csname natexlab\endcsname\relax\def\natexlab#1{#1}\fi
	
	\bibitem[{Abe(1993)}]{Abe2}
	Abe, Y. 1993, in Evolution of the Earth and Planets, ed. E.~Takahashi,
	R.~Jeanloz, \& D.~Rubie, Vol.~74 (AGU,{W}ashington {D.C.}), 41--53
	
	\bibitem[{{Bell} {et~al.}(2017){Bell}, {Murphy}, \& {Mamajek}}]{Bell2}
	{Bell}, C.~P.~M., {Murphy}, S.~J., \& {Mamajek}, E.~E. 2017, Mon. Not. R.
	Astron. Soc., 468, 1198
	
	\bibitem[{{Benz} \& {Cameron}(1990)}]{Benz}
	{Benz}, W. \& {Cameron}, A.~G.~W. 1990, Terrestrial effects of the Giant
	Impact., ed. H.~E. {Newsom} \& J.~H. {Jones}, 61--67
	
	\bibitem[{Bower {et~al.}(2018)Bower, Sanan, \& Wolf}]{Bowerm}
	Bower, D.~J., Sanan, P., \& Wolf, A.~S. 2018, Phys. Earth Planet. Inter., 274,
	49
	
	\bibitem[{Brandl {et~al.}(2016)Brandl, Agócs, Aitink-Kroes, Bertram,
		Bettonvil, van Boekel, Boulade, Feldt, Glasse, \& et~al.}]{Quanz2}
	Brandl, B., Agócs, T., Aitink-Kroes, G., {et~al.} 2016, Proc. SPIE, 9908
	
	\bibitem[{{Brandl} {et~al.}(2018){Brandl}, {Absil}, {Ag{\'o}cs}, {Baccichet},
		{Bertram}, {Bettonvil}, {van Boekel}, {Burtscher}, {van Dishoeck}, {Feldt},
		{Garcia}, {Glasse}, {Glauser}, {G{\"u}del}, {Haupt}, {Kenworthy}, {Labadie},
		{Laun}, {Lesman}, {Pantin}, {Quanz}, {Snellen}, {Siebenmorgen}, \& {van
			Winckel}}]{brandl2018}
	{Brandl}, B.~R., {Absil}, O., {Ag{\'o}cs}, T., {et~al.} 2018, in Society of
	Photo-Optical Instrumentation Engineers (SPIE) Conference Series, Vol. 10702,
	Society of Photo-Optical Instrumentation Engineers (SPIE) Conference Series,
	107021U
	
	\bibitem[{Canup \& Asphaug(2001)}]{Canup}
	Canup, R.~M. \& Asphaug, E. 2001, Nature, 412, 708
	
	\bibitem[{{Catling} \& {Kasting}(2017)}]{Catling&Kasting}
	{Catling}, D.~C. \& {Kasting}, J.~F. 2017, {Atmospheric Evolution on Inhabited
		and Lifeless Worlds, Cambridge University Press}
	
	\bibitem[{Cockell {et~al.}(2009)Cockell, Herbst, Léger, Absil, Beichmann,
		Benz, Brack, Chazelas, Chelli, Cottin, du~Foresto, Danchi, Defrère, den
		Herder, \& Eiroa}]{Darwin}
	Cockell, C., Herbst, T., Léger, A., {et~al.} 2009, Exp. Astron., 23
	
	\bibitem[{Costa {et~al.}(2009)Costa, Caricchi, \& Bagdassarov}]{Costa}
	Costa, A., Caricchi, L., \& Bagdassarov, N. 2009, Geochemistry Geophysics
	Geosystems, 10
	
	\bibitem[{{Defr{\`e}re} {et~al.}(2018){Defr{\`e}re}, {L{\'e}ger}, {Absil},
		{Garcia Munoz}, {Grenfell}, {Godolt}, {Loicq}, {Kammerer}, {Quanz}, {Rauer},
		{Schifano}, \& {Tian}}]{2018arXiv180709996D}
	{Defr{\`e}re}, D., {L{\'e}ger}, A., {Absil}, O., {et~al.} 2018, Proc. SPIE
	Astronomical Telescopes + Instrumentation 2018 (Austin, Texas), Optical and
	Infrared Interferometry and Imaging VI [\eprint[arXiv]{1807.09996}]
	
	\bibitem[{Demory {et~al.}(2016)Demory, Gillon, de~Wit, Madhusudhan, Bolmont,
		Heng, Kataria, Lewis, Hu, Krick, Stamenkovic, Benneke, Kane, \&
		Queloz}]{Demory}
	Demory, B.~O., Gillon, M., de~Wit, J., {et~al.} 2016, Nature, 532, 207
	
	\bibitem[{Elkins-Tanton(2012)}]{Elkins}
	Elkins-Tanton, L. 2012, Annu. Rev. Earth Planet. Sci., 40, 113
	
	\bibitem[{{Elkins-Tanton}(2008)}]{ET08}
	{Elkins-Tanton}, L.~T. 2008, Earth Planet. Sci. Lett., 271, 181
	
	\bibitem[{Elkins-Tanton {et~al.}(2011)Elkins-Tanton, Burgess, \&
		Yin}]{Tanton_Moon}
	Elkins-Tanton, L.~T., Burgess, S., \& Yin, Q. 2011, Earth Planet. Sci. Lett.,
	304, 326
	
	\bibitem[{Flasar \& Birch(1973)}]{Flasar}
	Flasar, F.~M. \& Birch, F. 1973, J. Geophys. Res., 78, 6101
	
	\bibitem[{Gagné {et~al.}(2017)Gagné, Faherty, Mamajek, Malo, Doyon,
		Filippazzo, Weinberger, Donaldson, Lépine, Lafrenière, Étienne Artigau,
		Burgasser, Looper, Boucher, Beletsky, Camnasio, Brunette, \& Arboit}]{Gagne}
	Gagné, J., Faherty, J.~K., Mamajek, E.~E., {et~al.} 2017, Astrophys. J.
	Suppl., 228, 18
	
	\bibitem[{Glasse {et~al.}(2015)Glasse, Rieke, Garcia-Marin, Ressler, Rost,
		Tikkanen, \& Vandenbussche}]{James}
	Glasse, A., Rieke, G., Garcia-Marin, M., {et~al.} 2015, Publ. Astron. Soc.
	Pac., 127
	
	\bibitem[{Grimm \& Stadel(2014)}]{Genga}
	Grimm, S.~L. \& Stadel, J.~G. 2014, Astrophys. J., 796
	
	\bibitem[{Hamano {et~al.}(2013)Hamano, Abe, \& Genda}]{Hamanonat}
	Hamano, K., Abe, Y., \& Genda, H. 2013, Nature, 497, 607
	
	\bibitem[{Hamano {et~al.}(2015)Hamano, Kawahara, Abe, Onishi, \&
		Hashimoto}]{Hamano}
	Hamano, K., Kawahara, H., Abe, Y., Onishi, M., \& Hashimoto, G.~L. 2015,
	Astrophys. J., 806
	
	\bibitem[{{Hammond} \& {Pierrehumbert}(2017)}]{Hammond17}
	{Hammond}, M. \& {Pierrehumbert}, R.~T. 2017, Astrophys. J., 849, 152
	
	\bibitem[{{Hartmann} \& {Davis}(1975)}]{1975Icar...24..504H}
	{Hartmann}, W.~K. \& {Davis}, D.~R. 1975, Icarus, 24, 504
	
	\bibitem[{{Hoffmann} {et~al.}(2017){Hoffmann}, {Grimm}, {Moore}, \&
		{Stadel}}]{Hoffmann17}
	{Hoffmann}, V., {Grimm}, S.~L., {Moore}, B., \& {Stadel}, J. 2017, Mon. Not. R.
	Astron. Soc., 465, 2170
	
	\bibitem[{Hunter(2007)}]{matplotlib}
	Hunter, J.~D. 2007, Comput. Sci. \& Eng., 9, 90
	
	\bibitem[{{Ikoma} {et~al.}(2018){Ikoma}, {Elkins-Tanton}, {Hamano}, \&
		{Suckale}}]{2018SSRv..214...76I}
	{Ikoma}, M., {Elkins-Tanton}, L., {Hamano}, K., \& {Suckale}, J. 2018, \ssr,
	214, 76
	
	\bibitem[{{Kammerer} \& {Quanz}(2018)}]{Kammerer}
	{Kammerer}, J. \& {Quanz}, S.~P. 2018, Astron. Astrophys., 609, A4
	
	\bibitem[{{Kasper} {et~al.}(2010){Kasper}, {Beuzit}, {Verinaud}, {Gratton},
		{Kerber}, {Yaitskova}, {Boccaletti}, {Thatte}, {Schmid}, {Keller}, {Baudoz},
		{Abe}, {Aller-Carpentier}, {Antichi}, {Bonavita}, {Dohlen}, {Fedrigo},
		{Hanenburg}, {Hubin}, {Jager}, {Korkiakoski}, {Martinez}, {Mesa}, {Preis},
		{Rabou}, {Roelfsema}, {Salter}, {Tecza}, \& {Venema}}]{kasper2010}
	{Kasper}, M., {Beuzit}, J.-L., {Verinaud}, C., {et~al.} 2010, in \procspie,
	Vol. 7735, Ground-based and Airborne Instrumentation for Astronomy III,
	77352E--77352E--9
	
	\bibitem[{Kokubo {et~al.}(2006)Kokubo, Kominami, \& Ida}]{Kokubo}
	Kokubo, E., Kominami, J., \& Ida, S. 2006, Astrophys. J., 642, 1131
	
	\bibitem[{Kraus {et~al.}(2014)Kraus, Shkolnik, Allers, \& Liu}]{Kraus}
	Kraus, A.~L., Shkolnik, E.~L., Allers, K.~N., \& Liu, C. 2014, Astrophys. J.
	
	\bibitem[{Lebrun {et~al.}(2013)Lebrun, Massol, E.Chassefière, Davaille, Marcq,
		Sarda, Leblanc, \& Brandeis}]{Lebrun}
	Lebrun, T., Massol, H., E.Chassefière, {et~al.} 2013, J. Geophys. Res.
	Planets, 118, 1155
	
	\bibitem[{Leger \& Herbst(2007)}]{Leger}
	Leger, A. \& Herbst, T. 2007, ArXiv e-prints [arXiv:0707.3385]
	
	\bibitem[{{Leinhardt} \& {Stewart}(2012)}]{LeinhardtStewartI}
	{Leinhardt}, Z.~M. \& {Stewart}, S.~T. 2012, Astrophys. J., 745, 79
	
	\bibitem[{Lupu {et~al.}(2014)Lupu, Zahnle, Marley, Schaefer, Fegley, Morley,
		Cahoy, Freedman, \& Fortney}]{Lupu}
	Lupu, R., Zahnle, K., Marley, M., {et~al.} 2014, Astrophys. J., 784
	
	\bibitem[{{Mamajek}(2016)}]{Mamajek}
	{Mamajek}, E.~E. 2016, arXiv:1507.06697, IAU Symposium, 314, 21
	
	\bibitem[{Mamajek \& Bell(2014)}]{Beta}
	Mamajek, E.~E. \& Bell, C. P.~M. 2014, MNRAS, 1
	
	\bibitem[{{Mamajek} \& {Meyer}(2007)}]{Mamajek07}
	{Mamajek}, E.~E. \& {Meyer}, M.~R. 2007, Astrophys. J. Lett., 668, L175
	
	\bibitem[{Marcq {et~al.}(2017)Marcq, Salvador, Massol, \& Davaille}]{Marcq17}
	Marcq, E., Salvador, A., Massol, H., \& Davaille, A. 2017, J. Geophys. Res.
	Planets, 122, 1539
	
	\bibitem[{Massol {et~al.}(2016)Massol, Hamano, Tian, Ikoma, Abe, Chassefière,
		Davaille, Genda, Güdel, Hori, Leblanc, Marcq, Sarda, Shematovich, Stökl, \&
		Lammer}]{Massol}
	Massol, H., Hamano, K., Tian, F., {et~al.} 2016, Space Sci. Rev., 205
	
	\bibitem[{Matsui \& Abe(1986)}]{Matsui2}
	Matsui, T. \& Abe, Y. 1986, Nature, 319, 303
	
	\bibitem[{{Melosh}(1990)}]{Melosh90}
	{Melosh}, H.~J. 1990, {Giant impacts and the thermal state of the early
		Earth.}, ed. N.~Y. O.~U. Press, 69--83
	
	\bibitem[{Miller-Ricci {et~al.}(2009)Miller-Ricci, Meyer, Seager, \&
		Elkins-Tanton}]{MillerRicci}
	Miller-Ricci, E., Meyer, M., Seager, S., \& Elkins-Tanton, L. 2009, Astrophys.
	J., 704, 770
	
	\bibitem[{Nakajima \& Stevenson(2015)}]{Naka}
	Nakajima, M. \& Stevenson, D. 2015, Earth Planet. Sci. Lett., 427, 286
	
	\bibitem[{Pujol \& North(2003)}]{Pujol}
	Pujol, T. \& North, G.~R. 2003, Tellus A: Dynamic Meteorology and Oceanography,
	55, 328
	
	\bibitem[{{Quanz} {et~al.}(2015){Quanz}, {Crossfield}, {Meyer}, {Schmalzl}, \&
		{Held}}]{Quanz}
	{Quanz}, S.~P., {Crossfield}, I., {Meyer}, M.~R., {Schmalzl}, E., \& {Held}, J.
	2015, Int. J. Astrobiol., 14, 279
	
	\bibitem[{{Quanz} {et~al.}(2018){Quanz}, {Kammerer}, {Defr{\`e}re}, {Absil},
		{Glauser}, \& {Kitzmann}}]{2018arXiv180706088Q}
	{Quanz}, S.~P., {Kammerer}, J., {Defr{\`e}re}, D., {et~al.} 2018, Proc. SPIE
	Astronomical Telescopes + Instrumentation 2018 (Austin, Texas), Optical and
	Infrared Interferometry and Imaging VI [\eprint[arXiv]{1807.06088}]
	
	\bibitem[{{Quintana} {et~al.}(2016){Quintana}, {Barclay}, {Borucki}, {Rowe}, \&
		{Chambers}}]{Quintana}
	{Quintana}, E.~V., {Barclay}, T., {Borucki}, W.~J., {Rowe}, J.~F., \&
	{Chambers}, J.~E. 2016, Astrophys. J., 821, 126
	
	\bibitem[{Raymond {et~al.}(2007)Raymond, Scalo, \& Meadows}]{Raymond}
	Raymond, S.~N., Scalo, J., \& Meadows, V.~S. 2007, Astrophys. J., 669, 606
	
	\bibitem[{Reese \& Solomatov(2006)}]{Reese}
	Reese, C.~C. \& Solomatov, V.~S. 2006, Icarus, 184, 102
	
	\bibitem[{Sasaki \& Nakazawa(1986)}]{Sasaki}
	Sasaki, S. \& Nakazawa, K. 1986, J. Geophys. Res. Solid Earth, 91, 9231
	
	\bibitem[{Shvonski {et~al.}(2016)Shvonski, Mamajek, Kim, Meyer, \& Pecaut}]{Sh}
	Shvonski, A.~J., Mamajek, E.~E., Kim, J.~S., Meyer, M.~R., \& Pecaut, M.~J.
	2016, ArXiv e-prints [arXiv:1612.06924]
	
	\bibitem[{Solomatov(2000)}]{Solomatov}
	Solomatov, V.~S. 2000, in Origin of the Earth and Moon, ed. R.~M. Canup \&
	K.~Righter, Space Science Series (The University of Arizona Press), 323--338
	
	\bibitem[{Solomatov \& Stevenson(1993)}]{Sol93}
	Solomatov, V.~S. \& Stevenson, D.~J. 1993, J. Geophys. Res., 98, 5375
	
	\bibitem[{{Stern}(1994)}]{Stern94}
	{Stern}, S.~A. 1994, Astron. J., 108, 2312
	
	\bibitem[{{The Astropy Collaboration} {et~al.}(2018){The Astropy
			Collaboration}, {Price-Whelan}, {Sip{\H o}cz}, {G{\"u}nther}, {Lim},
		{Crawford}, {Conseil}, {Shupe}, {Craig}, {Dencheva}, {Ginsburg},
		{VanderPlas}, {Bradley}, {P{\'e}rez-Su{\'a}rez}, {de Val-Borro}, {Aldcroft},
		{Cruz}, {Robitaille}, {Tollerud}, {Ardelean}, {Babej}, {Bachetti}, {Bakanov},
		{Bamford}, {Barentsen}, {Barmby}, {Baumbach}, {Berry}, {Biscani}, {Boquien},
		{Bostroem}, {Bouma}, {Brammer}, {Bray}, {Breytenbach}, {Buddelmeijer},
		{Burke}, {Calderone}, {Cano Rodr{\'{\i}}guez}, {Cara}, {Cardoso},
		{Cheedella}, {Copin}, {Crichton}, {D{\'A}vella}, {Deil}, {Depagne},
		{Dietrich}, {Donath}, {Droettboom}, {Earl}, {Erben}, {Fabbro}, {Ferreira},
		{Finethy}, {Fox}, {Garrison}, {Gibbons}, {Goldstein}, {Gommers}, {Greco},
		{Greenfield}, {Groener}, {Grollier}, {Hagen}, {Hirst}, {Homeier}, {Horton},
		{Hosseinzadeh}, {Hu}, {Hunkeler}, {Ivezi{\'c}}, {Jain}, {Jenness}, {Kanarek},
		{Kendrew}, {Kern}, {Kerzendorf}, {Khvalko}, {King}, {Kirkby}, {Kulkarni},
		{Kumar}, {Lee}, {Lenz}, {Littlefair}, {Ma}, {Macleod}, {Mastropietro},
		{McCully}, {Montagnac}, {Morris}, {Mueller}, {Mumford}, {Muna}, {Murphy},
		{Nelson}, {Nguyen}, {Ninan}, {N{\"o}the}, {Ogaz}, {Oh}, {Parejko}, {Parley},
		{Pascual}, {Patil}, {Patil}, {Plunkett}, {Prochaska}, {Rastogi}, {Reddy
			Janga}, {Sabater}, {Sakurikar}, {Seifert}, {Sherbert}, {Sherwood-Taylor},
		{Shih}, {Sick}, {Silbiger}, {Singanamalla}, {Singer}, {Sladen}, {Sooley},
		{Sornarajah}, {Streicher}, {Teuben}, {Thomas}, {Tremblay}, {Turner},
		{Terr{\'o}n}, {van Kerkwijk}, {de la Vega}, {Watkins}, {Weaver}, {Whitmore},
		{Woillez}, \& {Zabalza}}]{astropy18}
	{The Astropy Collaboration}, {Price-Whelan}, A.~M., {Sip{\H o}cz}, B.~M.,
	{et~al.} 2018, ArXiv e-prints [\eprint[arXiv]{1801.02634}]
	
	\bibitem[{Tonks \& Melosh(1993)}]{Tonks}
	Tonks, W. \& Melosh, H. 1993, Origin of the Earth, 98, 5319
	
	\bibitem[{Torres {et~al.}(2008)Torres, Quast, Melo, \& Sterzik}]{Torres}
	Torres, C.~A., Quast, G.~R., Melo, C. H.~F., \& Sterzik, M.~F. 2008, Handbook
	of Star Forming Regions, Astronomical Society of the Pacific, 2
	
	\bibitem[{{Wenger} {et~al.}(2000){Wenger}, {Ochsenbein}, {Egret}, {Dubois},
		{Bonnarel}, {Borde}, {Genova}, {Jasniewicz}, {Lalo{\"e}}, {Lesteven}, \&
		{Monier}}]{Simbad}
	{Wenger}, M., {Ochsenbein}, F., {Egret}, D., {et~al.} 2000, Astron. Astrophys.,
	143, 9
	
	\bibitem[{{Winn} \& {Fabrycky}(2015)}]{Winn15}
	{Winn}, J.~N. \& {Fabrycky}, D.~C. 2015, Annu. Rev. Astron. Astrophys., 53, 409
	
	\bibitem[{Wolf \& Bower(2018)}]{Wolf}
	Wolf, A. \& Bower, D. 2018, Phys. Earth Planet. Inter., 278, 59
	
	\bibitem[{Yamamoto \& Onishi(1952)}]{Yamamoto}
	Yamamoto, G. \& Onishi, G. 1952, Journal of Meteorology, 9, 415
	
	\bibitem[{Zahnle {et~al.}(1988)Zahnle, Kasting, \& Pollack}]{Zahnle88}
	Zahnle, K., Kasting, J., \& Pollack, J. 1988, Icarus, 74, 62
	
	\bibitem[{Zeng {et~al.}(2016)Zeng, Sasselov, \& Jacobsen}]{Mass}
	Zeng, L., Sasselov, D., \& Jacobsen, S. 2016, Astrophys. J., 819
	
	\bibitem[{{Zhang} \& {Sigurdsson}(2003)}]{Zhang03}
	{Zhang}, B. \& {Sigurdsson}, S. 2003, Astrophys. J. Lett., 596, L95
	
	\bibitem[{{Zuckerman} {et~al.}(2011){Zuckerman}, {Rhee}, {Song}, \&
		{Bessell}}]{ABDoradus}
	{Zuckerman}, B., {Rhee}, J.~H., {Song}, I., \& {Bessell}, M.~S. 2011,
	Astrophys. J., 732, 61
	
\end{thebibliography}

\noindent

\end{document}